\documentstyle[aaspp4]{article}
\def\pp{\par\parshape 2 0truecm 15.5truecm 1truecm 14.5truecm\noindent}
\newcommand{\simgt}{\lower.5ex\hbox{$\; \buildrel > \over \sim \;$}}
\newcommand{\simlt}{\lower.5ex\hbox{$\; \buildrel < \over \sim \;$}}
\newcommand{\lbkt}[1]{\left[#1\right]}

\newcommand{\sbkt}[1]{\left(#1\right)}
\newcommand{\abs}[1]{\left|#1\right|}
\newcommand{\msun}{\,{\rm M}_{\odot}}
\begin{document}

\title{Formation of Sub-galactic Clouds under UV Background Radiation}

\bigskip

\author{Tetsu Kitayama$^{1,2}$ and Satoru Ikeuchi$^{3}$}

\bigskip
\bigskip

\affil{
$^{1}$ Department of Physics, Tokyo Metropolitan University, 
Hachioji, Tokyo 192-0397, Japan\\
$^{2}$ Department of Physics, University of Tokyo, 
Bunkyo-ku, Tokyo 113-0033, Japan\\
$^{3}$ Department of Physics, Nagoya University, Chikusa-ku, Nagoya
464-8602, Japan \\
e-mail: tkita@phys.metro-u.ac.jp, ~
ikeuchi@a.phys.nagoya-u.ac.jp}

\received{1999 }
\accepted{  }

\begin{abstract}
The effects of the UV background radiation on the formation of
sub-galactic clouds are studied by means of one-dimensional
hydrodynamical simulations. The radiative transfer of the ionizing
photons due to the absorption by HI, HeI and HeII, neglecting the
emission, is explicitly taken into account.  We find that the complete
suppression of collapse occurs for the clouds with circular velocities
typically in the range $V_c \sim 15 - 40$ km s$^{-1}$ and the 50\%
reduction in the cooled gas mass with $V_c \sim 20 - 55$ km
s$^{-1}$. These values depend most sensitively on the collapse epoch
of the cloud, the shape of the UV spectrum, and the evolution of the
UV intensity. Compared to the optically thin case, previously
investigated by Thoul \& Weinberg (1996), the absorption of the
external UV photon by the intervening medium systematically lowers the
above threshold values by $\Delta V_c \sim 5$ km s$^{-1}$.  Whether
the gas can contract or keeps expanding is roughly determined by the
balance between the gravitational force and the thermal pressure
gradient when it is maximally exposed to the external UV flux. Based
on our simulation results, we discuss a number of implications on
galaxy formation, cosmic star formation history, and the observations
of quasar absorption lines. In Appendix, we derive analytical formulae
for the photoionization coefficients and heating rates, which
incorporate the frequency/direction-dependent transfer of external
photons.
\end{abstract}


\section{Introduction}

Photoionization of primordial gas is well known to have great impacts
on galaxy formation and the thermal history of the universe. The
observations of the QSO absorption spectra (Gunn \& Peterson 1965)
imply that the intergalactic gas has been highly ionized before
redshift $z \sim 5$. In fact, the existence of an intense ultraviolet
(UV) background radiation that can photoionize the universe is
inferred from the proximity effect of Ly$\alpha$ forest at $z = 2\sim
4$ (e.g. Baljtlik, Duncan \& Ostriker 1988; Bechtold 1994; Giallongo
et al. 1996). The origin of the UV background radiation could be
attributed to the emission from QSO's and/or young galaxies formed at
high redshifts (e.g. Couchman 1985; Miralda-Escude \& Ostriker 1990;
Fukugita \& Kawasaki 1994).

The UV background radiation, once produced, largely affects the
subsequent formation of structures. In particular, the formation of
low-mass objects is suppressed via photoionization and heating
associated with it (Umemura \& Ikeuchi 1985; Ikeuchi 1986; Rees 1986;
Couchman \& Rees 1986; Ikeuchi, Murakami \& Rees 1988, 1989;
Efstathiou 1992; Babul \& Rees 1992; Chiba \& Nath 1994; Babul \&
Ferguson 1996; Okoshi \& Ikeuchi 1996; Haiman, Rees, \& Loeb 1997;
Kepner, Babul \& Spergel 1997, among others). Such suppression may
partly resolve some of the shortcomings of the hierarchical models of
structure formation, such as an excess number of faint galaxies
predicted than actually observed (e.g. White \& Frenk 1991; Kauffmann,
White \& Guiderdoni 1993; Cole et al. 1994).  In view of this, several
authors have simulated the thermal and dynamical evolution of the
intergalactic medium under the UV background (Umemura \& Ikeuchi 1984;
Bond, Szalay \& Silk 1988; Murakami \& Ikeuchi 1990, 1993; Cen \&
Ostriker 1992; Zhang, Anninos \& Norman 1995; Quinn, Katz \&
Efstathiou 1996; Thoul \& Weinberg 1996; Weinberg, Hernquist \& Katz
1997; Navarro \& Steinmetz 1997).  For instance, Thoul \& Weinberg
(1996) concluded that the objects with circular velocities $V_c \simlt
30$ km s$^{-1}$ are prevented from collapsing at $2 \simlt z \simlt
5$. These previous studies, however, were based upon the assumption
that the medium is optically thin against the ionizing photons. Gas
clouds in reality become optically thick in the course of contraction
and it is by no means trivial how much the cloud evolution is altered
when the radiative transfer of ionizing photons is incorporated. In
addition, only the limited range of collapse redshift has been
explored previously.  As the mean density of the universe depends
strongly on redshift in proportion to $(1+z)^3$, the effects of the UV
background radiation should also vary, with an increasing importance
of the radiative transfer at higher redshifts.  The possible evolution
of the UV background intensity may have further impacts on the cloud
dynamics at different redshifts.

In this paper, we study the effects of the UV background radiation on
the dynamical evolution of primordial clouds, by means of
one-dimensional, spherically symmetric hydrodynamical simulations,
incorporating the radiative transfer of the ionizing photons. We use
analytical approximations of the photoionization coefficients and
heating rates derived in Appendix, which take explicit account of the
frequency/direction-dependent radiative transfer due to the absorption
by HI, HeI and HeII. They are simply expressed as a function of column
densities of each species from the cloud boundary and are applicable
to the medium with an arbitrary density profile. By using these
analytical formulae, we have achieved a significant reduction of the
computational time and are thus able to explore broad ranges in cloud
mass scale, collapse redshift, and parameters of the external UV
field. The results are compared quantitatively with those of optically
thin calculations. Based on the results of numerical simulations, we
further predict the global production rate of cooled gas in the
standard Cold Dark Matter (CDM) universe and discuss its implications
on galaxy formation and cosmic star formation history.  We also
discuss the imprints of the UV background on the quasar absorption
lines, such as the observability of the helium Ly$\alpha$ forests,
which can be tested by future observations.

The plan of this paper is as follows. In \S 2, we describe the
numerical model used in this paper. \S 3 presents the results of our
simulation and \S 4 is devoted to the discussion. Finally \S 4
summarizes our conclusions. Wherever necessary, the following
cosmological parameters are assumed for definiteness; the density
parameter $\Omega_0=1$, the Hubble constant $h=H_0/(100 \mbox{km
s$^{-1}$ Mpc$^{-1}$})=0.5$, the baryon density parameter $\Omega_b =
0.1$, and the amplitude of the density fluctuations
$\sigma_8=0.6$. The value of $\sigma_8$ is chosen to match the
observed local abundance of galaxy clusters (e.g. Viana \& Liddle
1996; Eke, Cole \& Frenk 1996; Kitayama \& Suto 1997).

\section{Method}

\subsection{Basic Equations}

We simulate the dynamics of a spherically symmetric bound system
exposed to the diffuse UV background radiation. The system is a
mixture of baryonic gas and collisionless dark matter, with the mass
ratio of $\Omega_b : \Omega_0 - \Omega_b = 1 : 9$.  The evolution of
these components is described by the following equations:
\begin{eqnarray}
\frac{dm_b}{dr_b} &=& 4 \pi r_b^2 \rho_b, \\
\frac{d^2r_b}{dt^2} &=& -4 \pi r_b^2 \frac{dP}{dm_b} -
\frac{G M(<r_b)}{r_b^2}, \\ 
\frac{du}{dt} &=& \frac{P}{\rho_b^2}\frac{d\rho_b}{dt} + \frac{{\cal H 
    - L}}{\rho_b}, \label{eq:energy}\\ 
P&=& (\gamma-1) \rho_b u = \frac{k_{\rm B}\rho_b T}{\mu m_{\rm p}}, 
\end{eqnarray}
and 
\begin{eqnarray}
\frac{d^2r_d}{dt^2} = - \frac{G M(<r_d)}{r_d^2}.
\end{eqnarray}
Here $r$, $m$, $\rho$, $P$, $T$, $u$, and $\mu$, are the radius, mass,
density, pressure, temperature, internal energy per unit mass, and
mean molecular weight in units of the proton mass $m_{\rm p}$,
respectively. $M(<r)$ is the total mass interior to $r$, ${\cal H}$
and ${\cal L}$ are the heating and cooling rates per unit volume, $G$
is the gravitational constant, and $k_{\rm B}$ is the Boltzmann
constant. Wherever necessary, the subscripts $b$ and $d$ denote baryon
and dark matter, respectively. The adiabatic index is fixed at
$\gamma=5/3$ throughout the paper.

The above equations are solved using the second-order-accurate
Lagrangian finite-difference scheme described in Bowers \& Wilson
(1991) and Thoul \& Weinberg (1995). The shocks are treated with the
artificial viscosity technique (Richtmyer \& Morton 1967; Umemura
1993).  The shells are binned equally in mass and their numbers are
$N_b=500$ for baryonic gas and $N_d= 5000$ for dark matter. We have
performed runs with $(N_b, N_d)=(2000, 20000)$ and $(300, 3000)$ and
confirmed that our results are robust against the changes in the
resolution. We have also checked that our code reproduces accurately
the similarity solutions for the adiabatic accretion of collisional
gas and for the pressure-less collapse onto a black hole (Bertschinger
1985). The total energy of the system is conserved by better than a
few percent in any runs reported in this paper.

\subsection{Radiative Processes}

The baryonic gas is assumed to have the primordial composition with
hydrogen and helium mass fractions $X=0.76$ and $Y=0.24$,
respectively. At each time-step, starting out from the cloud boundary
into the interior, we successively solve for the ionization
equilibrium among photoionization, collisional ionization and
recombination, together with the penetration of the external UV field,
as described in detail later in this section. We then compute the
heating/cooling rates due to photoionization, collisional
ionization/excitation, recombination, thermal bremsstrahlung, and
Compton scattering with the cosmic microwave background
radiation. Unless otherwise stated, we use the rates and coefficients
summarized in Fukugita \& Kawasaki (1994), which corrects a few typos
in Cen (1992).

The external UV field is taken to be isotropic and have the power-law
spectrum:
\begin{equation}
  J_{\rm in}(\nu,z) = J_{21}(z) \sbkt{\frac{\nu}{\nu_{\rm
        HI}}}^{-\alpha} \times 10^{-21} \mbox{erg s$^{-1}$ cm$^{-2}$
    str$^{-1}$ Hz$^{-1}$},
\label{eq:uvb}
\end{equation}
where $J_{21}(z)$ is the intensity (in proper coordinates) at the
Lyman limit of hydrogen ($h\nu_{\rm HI} = 13.6$eV) and $\alpha$ is the
spectral index. Observations of the proximity effect in the Ly$\alpha$
forest suggest $J_{21} = 10^{\pm 0.5}$ at $z=1.7-4.1$ (Baljtlik,
Duncan \& Ostriker 1988; Giallongo et al. 1996; Cooke, Espey \&
Carswell 1997; Savaglio et al. 1997), but its value is still highly
uncertain at other redshifts. Theoretical predictions of the value of
$J_{21}$ needed to reionize the universe at high redshifts range from
$J_{21} \sim 0.1$ to even $\sim 100$ (e.g. Fukugita \& Kawasaki 1994;
Gnedin \& Ostriker 1997; Haiman \& Loeb 1998; Madau, Haardt \& Rees
1999). The predicted epoch at which reionization occurs also has a
large uncertainty between $z \sim 50$ and 6. In the present paper,
unless otherwise stated explicitly, we fix the onset of the UV
background at $z_{\rm UV} =20$, and consider the following four cases
at $z<z_{\rm UV}$:
\begin{enumerate}
\item $J_{21} = 1$  and $\alpha=1$,
\item $J_{21} = 0.1$  and $\alpha=1$,
\item $J_{21} = 1$  and $\alpha=5$,
\item Evolving $J_{21}$ and $\alpha=1$, where \\
\begin{equation}
J_{21} = \left\{\begin{array}{ll}
      \sbkt{\frac{1+z}{7}}^{-6} &  6 \leq z \leq z_{\rm UV} \\
      1 &  3 \leq z \leq 6 \\ 
      \sbkt{\frac{1+z}{4}}^4 &  0 \leq z \leq 3. \\ 
     \end{array} \right.
\label{eq:uvevol}
\end{equation}
\end{enumerate}
The values $\alpha=1$ and $5$ are chosen to mimic the spectra of
quasars and massive stars, respectively. By the above form of the UV
evolution (eq.[\ref{eq:uvevol}]), we attempt to study the effects of a
late reionization and a decline of the UV intensity at low redshift.

The incident UV spectrum is modified due to the radiative transfer as
it penetrates into the gas cloud. In the present paper, we explicitly
take account of the frequency/direction-dependent absorption by HI,
HeI, and HeII, using equations (\ref{eq:coeff4}) and (\ref{eq:rate4})
in Appendix. Having performed the frequency and angular integrations
of the transfer equation neglecting the emission term, the analytical
formulae are obtained for the photoionization coefficients and heating
rates in the plane-parallel slab. They are simply expressed as a
function of column densities of individual species measured from the
boundary, and lead to a significant reduction in the computational
time. By means of these formulae, we solve simultaneously for the
ionization equilibrium in each gas shell and the UV radiation field
processed between the outer boundary and that shell. The ionization
state of the gas and the radiation field are computed iteratively
until the abundances of HI, HeI and HeII all converge to the precision
better than 1\%.

Note that the above treatment of the radiative transfer systematically
overestimates the effect of absorption in the following respects; 1)
since the solution of the transfer equation in the plane-parallel
geometry is applied to a spherical cloud, the photon path lengths from
the outer boundary are overestimated, 2) at each radial point,
ionizing photons coming from the inner $2\pi$ steradian of the cloud
is ignored, and 3) the emission (or scatter) of incident photons is
neglected.  It should thus provide a conservative limit in which an
external UV field is maximally weakened and has the minimal effects on
cloud evolution.  This is in fact complementary to a conventional
approximation of the optically thin medium (e.g.  Thoul \& Weinberg
1996), in which the external UV field is likely to have the maximal
effects on the cloud dynamics. In what follows, therefore, we examine
both of these limiting cases and perform quantitative comparisons
between them.  In so doing, we hope to bracket the true answer, which
is still very difficult to solve in a fully self-consistent manner.

The approximation of the ionization equilibrium is correct if the
recombination time-scale $t_{\rm rec}$ is shorter than the dynamical
time-scale $t_{\rm dyn} \equiv 1/\sqrt{G \rho}$. Assuming a high
degree of ionization, this yields the following condition for the
electron density $n_e$ (e.g. Vedel, Hellsten \& Sommer-Larsen 1994):
\begin{equation}
n_e > 7.1 \times 10^{-6} T_4^{1.4} \sbkt{\frac{2}{1+X}}
\sbkt{\frac{\Omega_0/\Omega_b}{10}} \mbox{ ~~cm$^{-3}$},
\end{equation}
where $T_4 \equiv (T/10^4 \mbox{ K})$, and we have adopted $t_{\rm
rec} \equiv (\alpha_{\rm H} n_e)^{-1}$ using the recombination rate
for hydrogen $\alpha_{\rm H}$ given in equation (\ref{eq:hrec}). The
above condition is satisfied in most situations considered in this
paper except at the outer envelopes of clouds at low redshifts. For
example, in a cloud collapsing at $z = 3$, typically the out-most
$\sim 5 \%$ of the gas has the density below the above value at $z=3$
(see \S \ref{sec:init} for the definition of the collapse epoch). For
these shells, the assumption of the ionization equilibrium will result
in overestimating the fractions of neutral species and hence in
overestimating the effect of absorption even further. In practice,
however, the physical abundances of neutral species in these shells
are negligibly small. Thus the approximation of the ionization
equilibrium is well justified in the present analysis.

\subsection{Initial Conditions}
\label{sec:init}

We start the simulations when the overdensity of a cloud is still in
the linear regime. The initial overdensity profile is assumed to have
the form of a single spherical Fourier mode for both baryon and dark
matter components:
\begin{equation}
\delta_{i}(r) = \delta_{i}(0) \frac{\sin(kr)}{kr}, 
\label{eq:deltai}
\end{equation}
where $k$ is the comoving wavenumber, and $\delta_{i}(0)$ is the
central overdensity.  Throughout the paper, we fix
$\delta_i(0)=0.2$. Assuming that the initial perturbation is dominated
by the growing mode, the initial velocity profile is given by
\begin{equation}
  v_{i}(r) = H_i r \sbkt{1 - \frac{\bar{\delta}_i(<r)}{3}},
\label{eq:vi}
\end{equation}
where $H_i$ is the Hubble parameter at the initial epoch, and
$\bar{\delta}(<r)$ is the volume averaged overdensity within radius
$r$. The outer boundary is taken at the first minimum of
$\delta_{i}(r)$, i.e. $kr = 4.4934$, at which $\bar{\delta}(<r)$
vanishes and the shell initially expands at the speed $H_i r$. As in
Haiman, Thoul \& Loeb (1996), we define the baryonic mass enclosed
within this radius as the bound mass $M_{\rm bound}$, and that
enclosed within the first zero of $\delta_{i}(r)$, i.e. $kr = \pi$, as
the cloud mass $M_{\rm cloud}$.  These masses are related to each
other via $M_{\rm cloud} = 0.342[1+0.304 \delta_i(0)] M_{\rm bound} =
0.363 M_{\rm bound}$.

Having fixed the initial density profile, we vary the initial redshift
$z_i$ and the cloud mass $M_{\rm cloud}$ to simulate different
collapse epochs and mass scales, i.e., circular velocities or virial
temperatures.  We define the central collapse redshift $z_{c0}$ and
the cloud collapse redshift $z_c$ respectively as the epochs at which
the inner-most gas shell and the shell enclosing $M_{\rm cloud}$ would
collapse to the center in the absence of thermal pressure.  The
circular velocity $V_c$ and the virial temperature $T_{\rm vir}$ of
the cloud are defined using $M_{\rm cloud}$ and $z_c$ as
\begin{eqnarray}
  V_c &=& 15.9 \sbkt{\frac{M_{\rm cloud} \Omega_0/\Omega_b}
{10^9 h^{-1} \msun}}^{1/3} (1+z_c)^{1/2} ~~~\mbox{  km s$^{-1}$},  \\
T_{\rm vir} &=& 9.09 \times 10^3  \sbkt{\frac{\mu}{0.59}}
\sbkt{\frac{M_{\rm cloud} \Omega_0/\Omega_b}
{10^9 h^{-1} \msun}}^{2/3} (1+z_c) ~~~\mbox{  K}.
\label{eq:vc}
\end{eqnarray}
Unless otherwise stated explicitly, we study three combinations of
$z_i$, $z_{c0}$, and $z_c$ listed in Table \ref{tab:zi}. In this
table, we also list the turn-around redshifts corresponding to
$z_{c0}$ and $z_c$, $z_{\rm ta0} \equiv 2^{2/3}(1+z_{c0})-1$ and
$z_{\rm ta} \equiv 2^{2/3}(1+z_c)-1$, respectively.  In the
low-redshift collapse, the gas is exposed to the UV background
radiation from the linear regime. In the middle-redshift collapse, the
cloud center is close to turn-around at the onset of the UV radiation
$z_{\rm UV}=20$. In the high-redshift collapse, the cloud center has
already collapsed prior to $z_{\rm UV}$.

At an initial epoch, the gas is assumed to have the uniform
temperature given by
\begin{equation}
  T_i = \left\{\begin{array}{ll} 2.726 (1+z_i) & ~~~~\mbox{ if $z_i
        \geq 200$} \nonumber \\ 548
      \sbkt{\frac{1+z_i}{201}}^{3(\gamma-1)} & ~~~~\mbox{ if $z_i <
        200$}, \\ 
     \end{array} \right.
\label{eq:ti}
\end{equation}
taking account of the fact that the matter is tightly coupled to the
cosmic microwave background at $z\simgt 200$ (e.g. Anninos \& Norman
1994; Chieze, Teyssier \& Alimi 1997).

\subsection{Boundary Conditions}

If the gas is able to lose sufficient energy by radiative cooling, it
falls towards the center nearly at the free-fall rate. As the density
increases near the center, the cooling time-scale $t_{\rm cool}\equiv u
\rho/\abs{{\cal H - L}}$ and the dynamical time-scale $t_{\rm dyn}$
become extremely small, and infinitely large number of time-steps are
required in the simulation. In order to avoid this, we introduce the
following criteria. If a gas shell reaches below some minimum radius
$r_{\rm min}$ {\it and} satisfy $t_{\rm cool}/t_{\rm dyn}< 0.01$, then
we regard it as having cooled and collapsed; we move the shell to the
center and ignore in the rest of the simulation, except in the
calculation of the gravitational force. When dropping a shell to the
center, we assume that the next shell expands inward adiabatically to
cover the volume of the dropped shell. The evolution of each shell is
traced until both of the above criteria are fulfilled.  We adopt as
$r_{\rm min} $ the radius at which the system attains rotational
support (e.g. Padmanabham 1993):
\begin{eqnarray}
r_{\rm min} &=& 0.05 \sbkt{\frac{\Omega_b/\Omega_0}{0.1}}^{-1} 
\sbkt{\frac{\lambda_{\rm ta}}{0.05}}^{2} r_{\rm ta}, 
\label{eq:barrier}
\end{eqnarray}
where $r_{\rm ta}$ is the turn-around radius of the gas shell, and
$\lambda_{\rm ta}$ is the dimensionless spin parameter attained by
turn-around. We fix $\lambda_{\rm ta}=0.05$ based on the results of
the numerical simulations (Efstathiou \& Jones 1979; Barns \&
Efstathiou 1987; Warren et al. 1992). Equation (\ref{eq:barrier}) also
gives the scale under which the assumption of spherical symmetry
breaks down. The baryon mass that has collapsed to the center is
denoted by $M_{\rm coll}$.

The dynamical time-scale also becomes very small when collisionless
dark matter shells approach the center. We therefore rebounce the dark
matter shells near the center (Spitzer \& Hart 1971; Gott 1975). As
noted by Thoul \& Weinberg (1995), it is desirable to choose the
rebouncing radius $r_{\rm reb}$ that is small enough not to alter the
evolution of gas shells, but not too small to degrade energy
conservation. In order to achieve these, we set $r_{\rm reb}$ equal to
$r_{\rm min}$ of a gas shell that encloses $0.01 - 0.1 M_{\rm bound}$
depending on a run. Throughout each run, $r_{\rm reb}$ is fixed at
this value. 

At the outer boundary, we adopt the free boundary condition, i.e. the
pressure outside the cloud $P_{\rm out}$ is zero. We have also
performed runs with a mirror boundary condition ($P_{\rm out}=
P_{N_b}$, where $P_{N_b}$ is the pressure of the outermost gas shell)
and confirmed that our results remain essentially invariant.
 
\section{Numerical Results}

\subsection{Dynamical evolution}
\label{sec:dyn}

\begin{figure}
\plotone{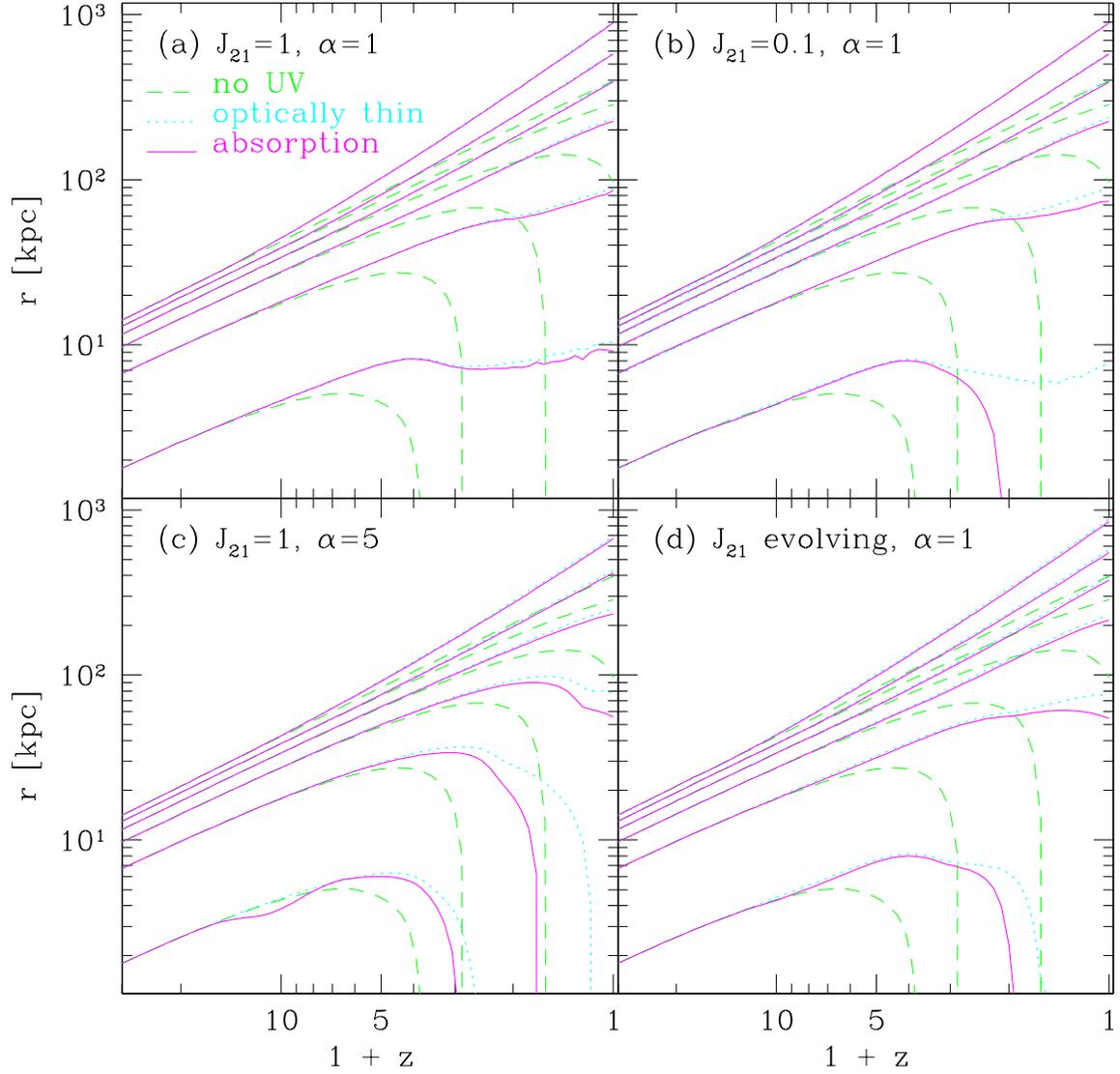} 
\caption{Trajectories of radii of gas shells enclosing 0.2 (inner-most
shell), 10, 30, 50, 70, 90 \% of $M_{\rm bound}$ in the low-redshift
collapse ($z_{c} = 0.5$) for a cloud with $V_c=32$ km s$^{-1}$
($M_{\rm cloud}\simeq 9 \times 10^8 \msun$) and different parameters
of the UV background; (a) $J_{21}=1$, $\alpha=1$, (b) $J_{21}=0.1$,
$\alpha=1$, (c) $J_{21}=1$, $\alpha=5$, and (d) evolving $J_{21}$,
$\alpha=1$. Different lines indicate the no UV case (dashed), the
optically thin case (dotted), and the case with absorption (solid).
\label{fig:r_z1}}
\end{figure}

Figure \ref{fig:r_z1} illustrates the trajectories of gas shell radii
in the low-redshift collapse ($z_c =0.5$) for a cloud with $V_c = 32$
km s$^{-1}$ ($M_{\rm cloud} \simeq 9 \times 10^8 \msun$). In the
absence of the UV background, the gas shells initially comoving with
the Hubble expansion will eventually turn around due to gravity and
start to contract. As the virial temperature of the cloud is $T_{\rm
vir} \simeq 4 \times 10^{4}$ K, the gas can lose energy efficiently by
radiative cooling and collapse towards the center. If the UV
background radiation is present, on the contrary, the evolution of the
cloud is significantly altered.  Figure \ref{fig:r_z1} (a) shows that
the cloud is prohibited to collapse completely under a constant UV
with $J_{21}=1$ and $\alpha=1$. Once, however, the UV amplitude is
reduced to $J_{21}=0.1$ and the absorption is taken into account, the
inner part of the cloud may start to collapse (Fig.~\ref{fig:r_z1}b).
Alternatively, a softer spectrum $\alpha=5$ enables larger number of
shells to collapse by $z=0$ (Fig.~\ref{fig:r_z1}c). The cloud collapse
is also promoted if the UV background radiation evolves according to
equation (\ref{eq:uvevol}) (Fig.~\ref{fig:r_z1}d).

\begin{figure}
\plotone{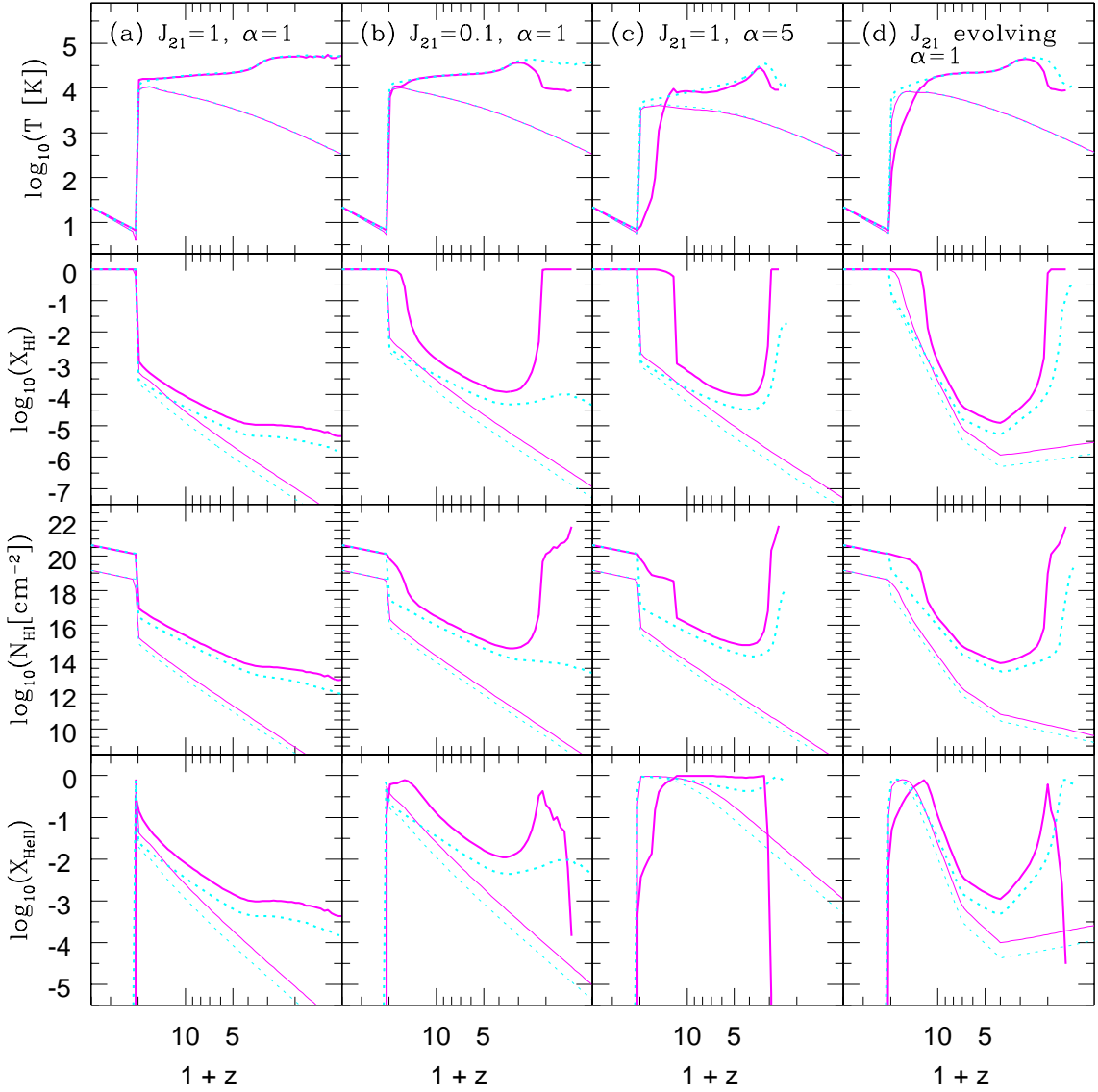} 
\caption{Evolution of temperature (highest panels), $X_{\rm HI}$
(second highest), $N_{\rm HI}$ (second lowest), and $X_{\rm HeII}$
(lowest) of gas shells enclosing 0.2\% (thick lines) and 90\% (thin
lines) of $M_{\rm bound}$ in the low-redshift collapse ($z_{c} = 0.5$)
for a cloud with $V_c=32$ km s$^{-1}$.  (a) $J_{21}=1$, $\alpha=1$,
(b) $J_{21}=0.1$, $\alpha=1$, (c) $J_{21}=1$, $\alpha=5$, and (d)
evolving $J_{21}$, $\alpha=1$. Different lines indicate the optically
thin case (dotted lines) and the case with absorption (solid lines).
\label{fig:txx_z1}}
\end{figure}

In order to study the above mentioned features in more detail, we plot
in Figure \ref{fig:txx_z1} the evolution of the temperature $T$, HI
fraction $X_{\rm HI} \equiv n_{\rm HI}/(n_{\rm HI}+n_{\rm HII})$, HI
column density measured from the outer boundary $N_{\rm HI}$, and HeII
fraction $X_{\rm HeII} \equiv n_{\rm HeII}/(n_{\rm HeI}+n_{\rm
HeII}+n_{\rm HeIII})$ in the runs shown in Figure \ref{fig:r_z1}.  For
$J_{21}=1$ and $\alpha=1$, temperature rises to $T \sim 10^4$ K and
all the gas is photoionized almost instantaneously at the onset of the
UV background $z_{\rm UV}=20$ (Fig.~\ref{fig:txx_z1}a). At this epoch,
the cloud density is still almost equal to the mean of the universe
($n_{\rm H} \sim 2 \times 10^{-3}$ cm$^{-3}$) and the external UV
field can penetrate into the cloud center, even if the absorption is
considered. The thermal pressure of the heated gas takes over the
gravity and the cloud keeps expanding.

For a smaller UV amplitude $J_{21}=0.1$, Figure \ref{fig:txx_z1}(b)
shows that the cloud evolves very differently with/without the
absorption. If the absorption is considered, the central part of the
cloud is not ionized instantaneously, but stays neutral until the gas
density decreases significantly due to expansion. It is noticeable
that the temperature still rises almost instantaneously to $T \sim
10^4$ K. This is because the time-scale for heating is much shorter
than that for ionization (e.g. Gnedin \& Ostriker 1997). One can
roughly estimate the ratio of these time-scales in the case of
$\alpha=1$ as
\begin{eqnarray}
\frac{t_{\rm heat}}{t_{\rm ion}} &=& \left.
\frac{3 k_{\rm B}T}{2 {\cal H}_{\rm HI}} \right/ \Gamma_{\rm HI}^{-1}
\nonumber \\
&=& \left\{\begin{array}{ll}
0.285 T_4 &  \mbox{ optically thin limit}, \\
7.28 \times 10^{-2} T_4  \tau_{\rm HI}^{-1/3} & 
\mbox{ optically thick limit},
     \end{array} \right.  
\label{eq:heation}
\end{eqnarray}
where $\tau_{\rm HI} = 6.3 \times 10^{-18} (N_{\rm
HI}/\mbox{cm}^{-2})$ is the HI optical depth at the Lyman limit
$\nu_{\rm HI}$, and $\Gamma_{\rm HI}$ and ${\cal H}_{\rm HI}$ are
photoionization coefficient and heating rate derived for the pure
hydrogen gas in Appendix (eqs~[\ref{eq:coeff6}]--[\ref{eq:rate7}]). At
$z_{\rm UV}$, the HI column density at the center is $N_{\rm HI} \sim
10^{20}$ cm$^{-2}$ in Figure \ref{fig:txx_z1}(b), which gives the
optical depth of $\tau_{\rm HI}=630$ and $t_{\rm heat}/t_{\rm ion}=
8.5 \times 10^{-3} T_4$ in the above equation. For a softer spectrum,
this ratio becomes larger and the heating becomes relatively slower as
illustrated in Figure \ref{fig:txx_z1}(c). In this case, the pressure
gradient between the cold neutral center of the cloud and the hot
ionized envelope causes to push the gas slightly inward near the cloud
center at $z \sim 10$ (Fig. \ref{fig:r_z1} c).

Figure \ref{fig:txx_z1}(b) further indicates that as the gravity
overcomes the pressure support, the gas shells can turn around and the
density starts to rise again. The gas is first heated by shock and
adiabatic compression, but it will start to lose energy and shrink to
the center once the radiative cooling becomes efficient. Consequently
the neutral core is formed at the center. The absorption of the
external UV field promotes the cloud collapse and the formation of the
neutral core, by both reducing the number of photoionized ions and
lowering the heating rate.

The cloud evolution is very sensitive to the spectral index of the UV
background. For a softer spectrum $\alpha=5$, there are smaller
numbers of high energy photons, and hence larger numbers of HeII, than
for $\alpha=1$ (Fig.~\ref{fig:txx_z1}c). The larger number of shells
are thus able to collapse within a given time for $\alpha=5$.

If the UV background evolves according to equation (\ref{eq:uvevol}),
the gas in the cloud is ionized gradually as in Figure
\ref{fig:txx_z1}(d). As the UV intensity drops at $z < 3$, the neutral
fraction starts to rise again and the inner shells start to collapse
to the center.

\begin{figure}
\plotone{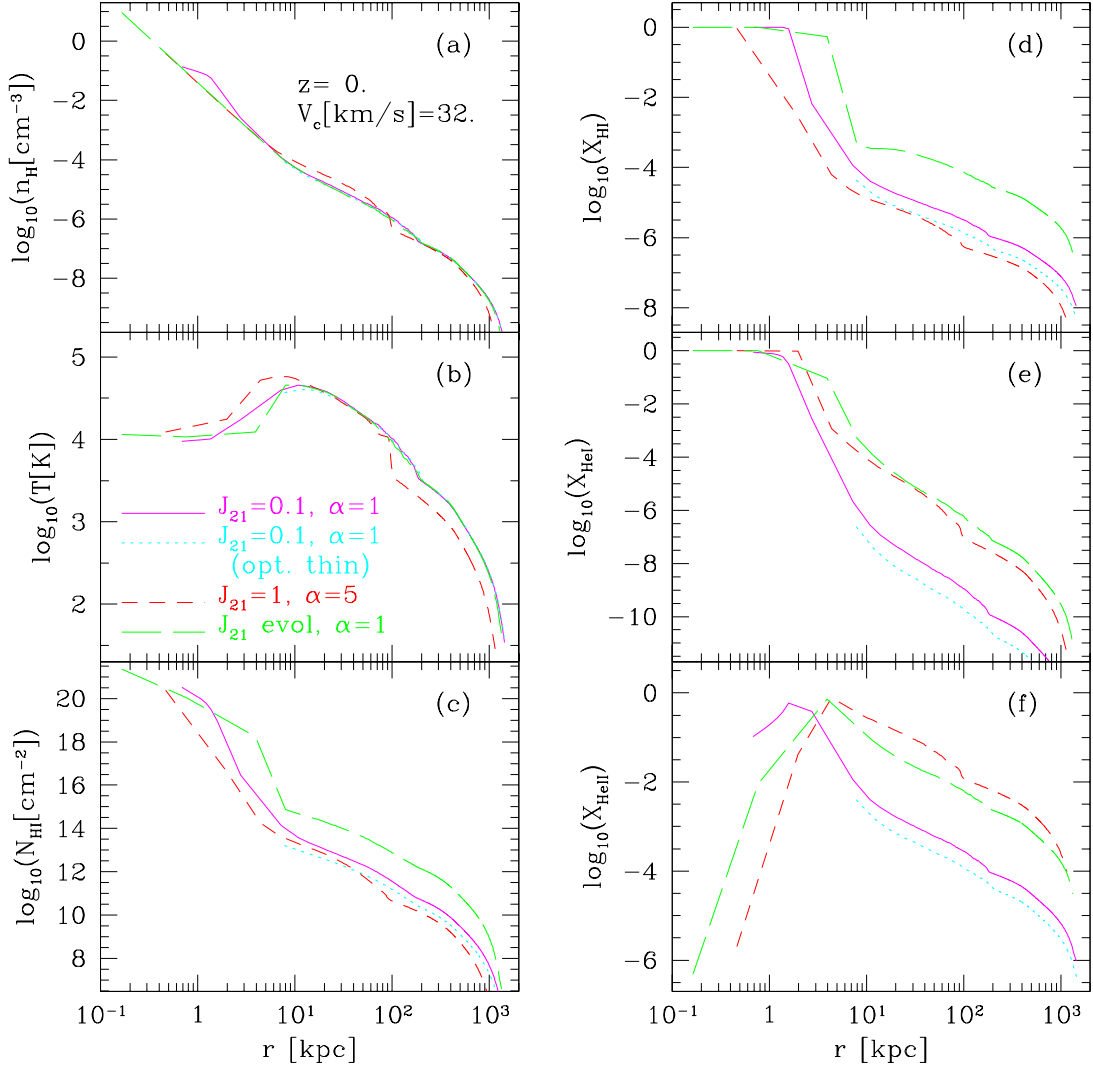}
\caption{Radial profiles at $z=0$ of (a) hydrogen density, (b)
  temperature, (c) HI column density from the boundary, (d) $X_{\rm
  HI}$, (e) $X_{\rm HeI}$, and (f) $X_{\rm HeII}$ in the low-redshift
  collapse ($z_c=0.5$) for a cloud with $V_c=32$ km s$^{-1}$. Lines
  correspond to different parameters of the UV background as shown in
  the figure (unless indicated explicitly, absorption is taken into
  account).
\label{fig:prof}}
\end{figure}

Figure \ref{fig:prof} shows the radial profiles at $z=0$ in some of
the runs discussed above. The shock front (if exists) is at $r \sim
10^2$ kpc.  The density profile inside the shock front agrees well
with the self-similar solution $\rho \propto r^{-2.25}$ (Bertschinger
1985). The central part of the cloud is self-shielded against the
external UV photons and the gas becomes neutral. In the case of
$(J_{21},\alpha) = (0.1,1)$ with absorption, the simulated profiles of
$X_{\rm HI}$ and $N_{\rm HI}$ roughly agree with an analytical
estimation for the pure hydrogen gas presented in Appendix.  For
instance, inserting into equation (\ref{eq:depth2}) the values read
off from the figure at $r \sim 2$ kpc, ($n_{\rm H} \sim
10^{-2}$cm$^{-2}$, $T \sim 10^4$K, and $X_{\rm HI} \sim 0.1$), we
obtain $\tau_{\rm HI} \sim 2$, i.e.  $N_{\rm HI} \sim 3 \times
10^{17}$cm$^{-2}$, in good accordance with Figure \ref{fig:prof}(c).
On the contrary, such an estimation breaks down for the softer
spectrum ($\alpha=5$) because of the larger abundances of HeI and
HeII.

\begin{figure}
\plotone{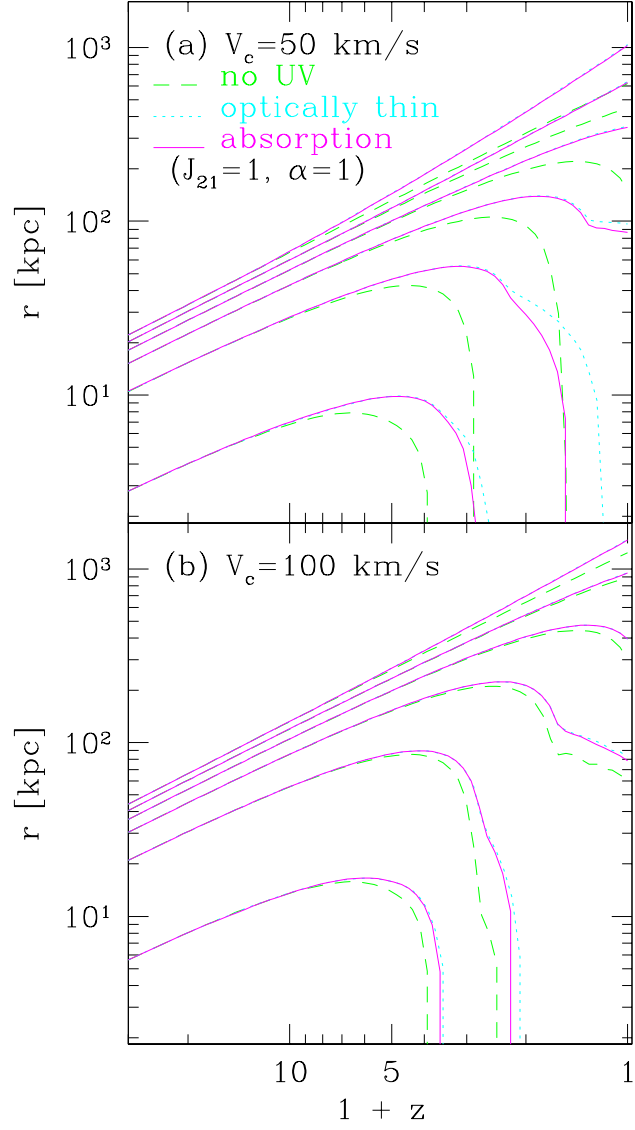}
\caption{Same as Fig.~\protect\ref{fig:r_z1}, except for a cloud with
  (a) $V_c=50$ km s$^{-1}$ ($M_{\rm cloud}\simeq 3 \times 10^{9}
  \msun$), and (b) $V_c=100$ km s$^{-1}$ ($M_{\rm cloud}\simeq 3
  \times 10^{10} \msun$), in the case of $J_{21}=1$ and $\alpha=1$.
\label{fig:r_z1b}}
\end{figure}

The effects of the UV background radiation become less significant as
the cloud mass increases. Figure \ref{fig:r_z1b}(a) shows that the gas
collapse is still delayed by the UV radiation for a cloud with $V_c =
50$ km s$^{-1}$ ($M_{\rm cloud}\simeq 3 \times 10^{9} \msun$, $T_{\rm
vir} \simeq 9 \times 10^4$ K). The evolution, however, is only
slightly altered for a larger cloud with $V_c = 100$ km s$^{-1}$
($M_{\rm cloud}\simeq 3 \times 10^{10} \msun$, $T_{\rm vir} \simeq 4 \times
10^5$ K). The effects of absorption, compared to the optically thin
case, also become insignificant at this mass scale as the collisional
ionization dominates the photoionization.

\begin{figure}
\plotone{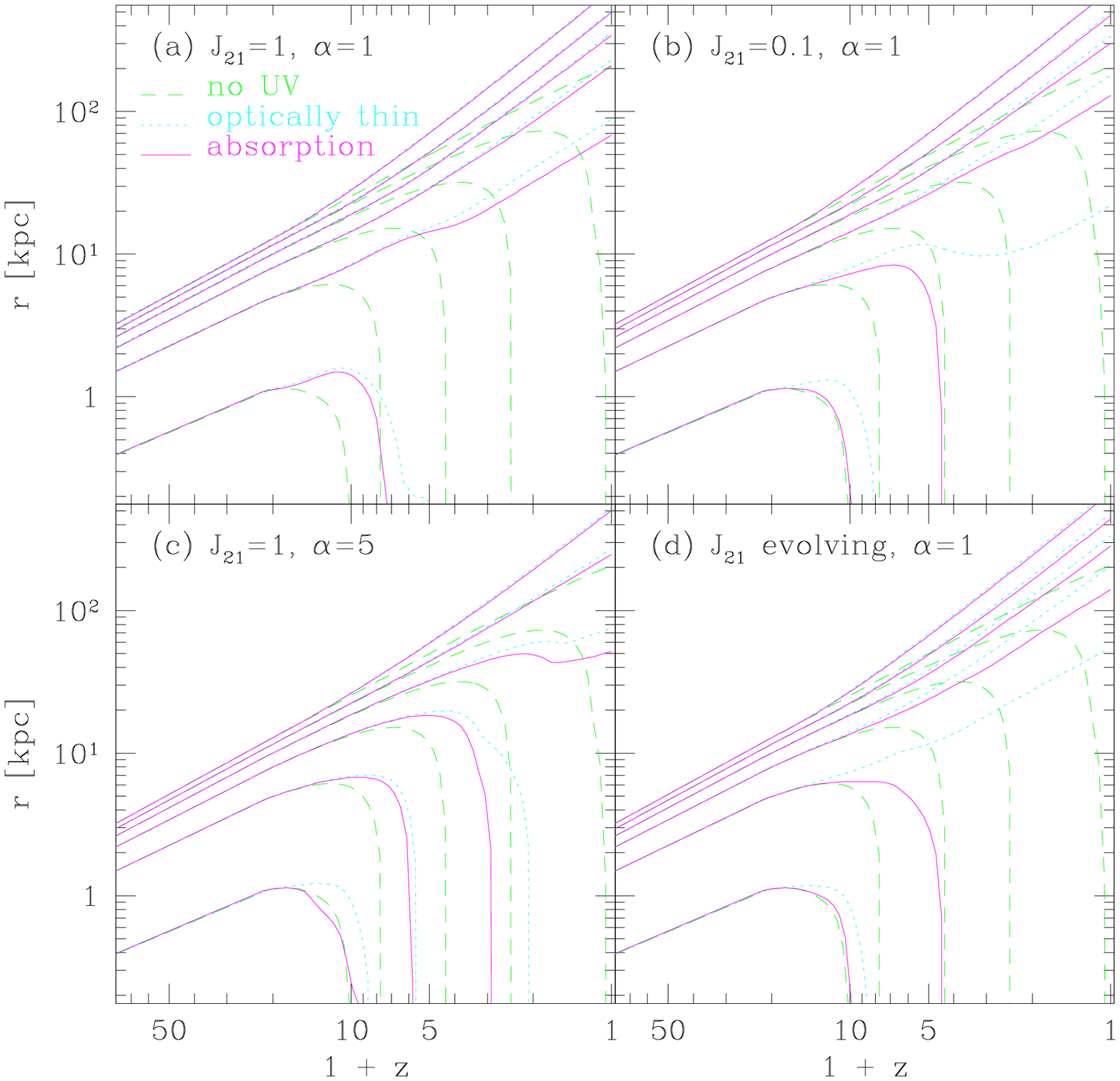}
\caption{Same as Fig.~\protect\ref{fig:r_z1}\protect, except for the
  middle-redshift collapse ($z_{c} = 3$) with $V_c=32$ km s$^{-1}$
  ($M_{\rm cloud} \simeq 2 \times 10^8 \msun$).
\label{fig:r_z2}}
\end{figure}

\begin{figure}
\plotone{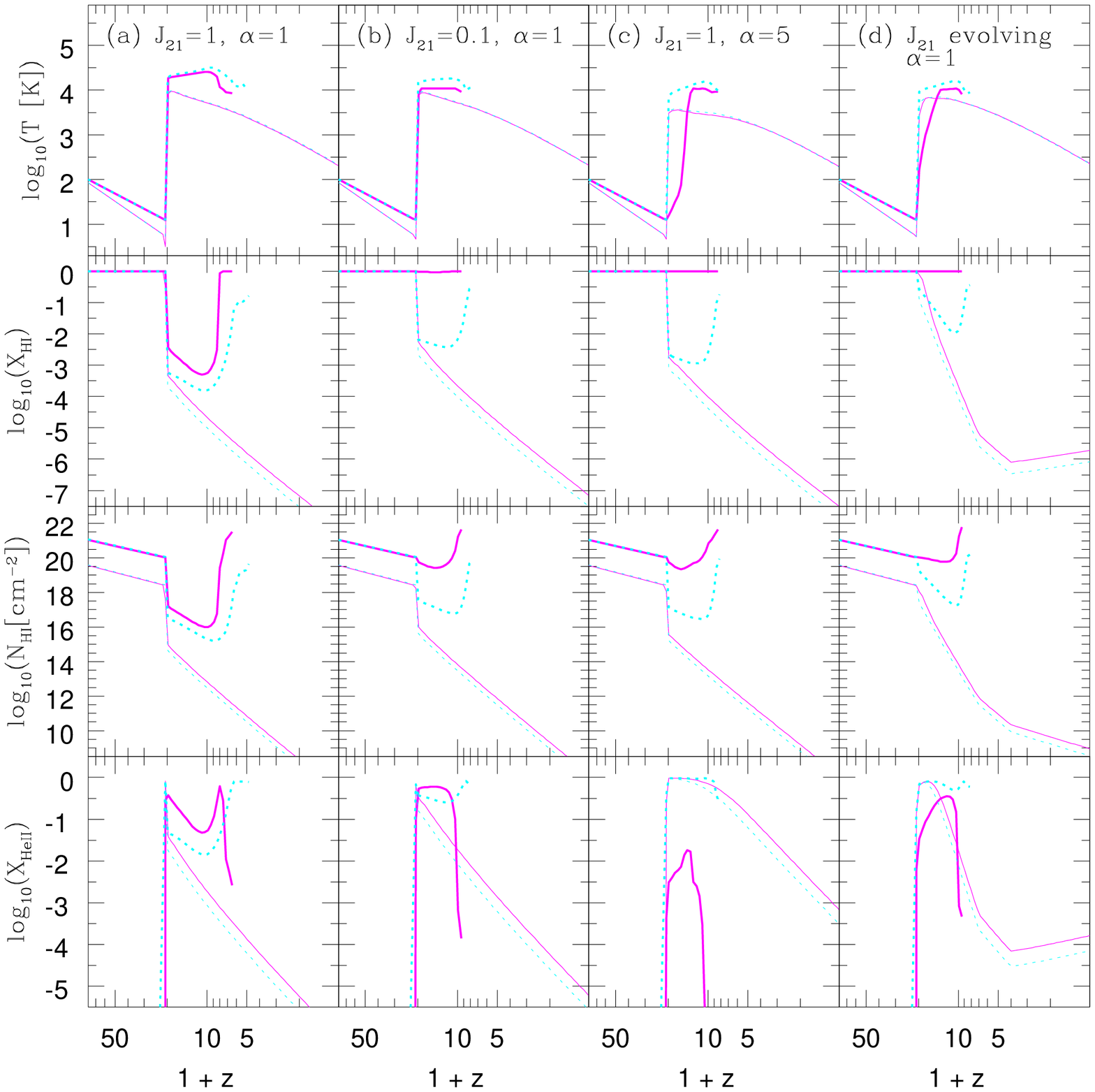}
\caption{Same as Fig.~\protect\ref{fig:txx_z1}\protect, except for the
  middle-redshift collapse ($z_{c} = 3$) with $V_c=32$ km s$^{-1}$
  ($M_{\rm cloud} \simeq 2 \times 10^8 \msun$).
\label{fig:txx_z2}}
\end{figure}

Figures \ref{fig:r_z2} and \ref{fig:txx_z2} illustrate the
trajectories of gas shells in the middle-redshift collapse ($z_c =
3$). Since the density of a cloud is higher than the low-redshift
collapses discussed above, the larger amount of gas is able to
collapse for a given circular velocity $V_c=32$ km s$^{-1}$. At the
onset of the UV background $z_{\rm UV}=20$, the cloud center is about
to turn around, and the density is $n_{\rm H} \sim 10^{-2}$ cm$^{-3}$,
roughly a factor of 5 larger than the mean density of the Universe. At
this density, the inner part of the cloud can be kept self-shielded
against ionizing photons until it collapses if the absorption is
considered (Fig.~\ref{fig:txx_z2}b,c,d). Even in such cases, the gas
is still heated promptly to $T\sim 10^4$K for $\alpha=1$, because of
the high efficiency of heating process over ionization mentioned above
(eq.~[\ref{eq:heation}]).

\begin{figure}
\plotone{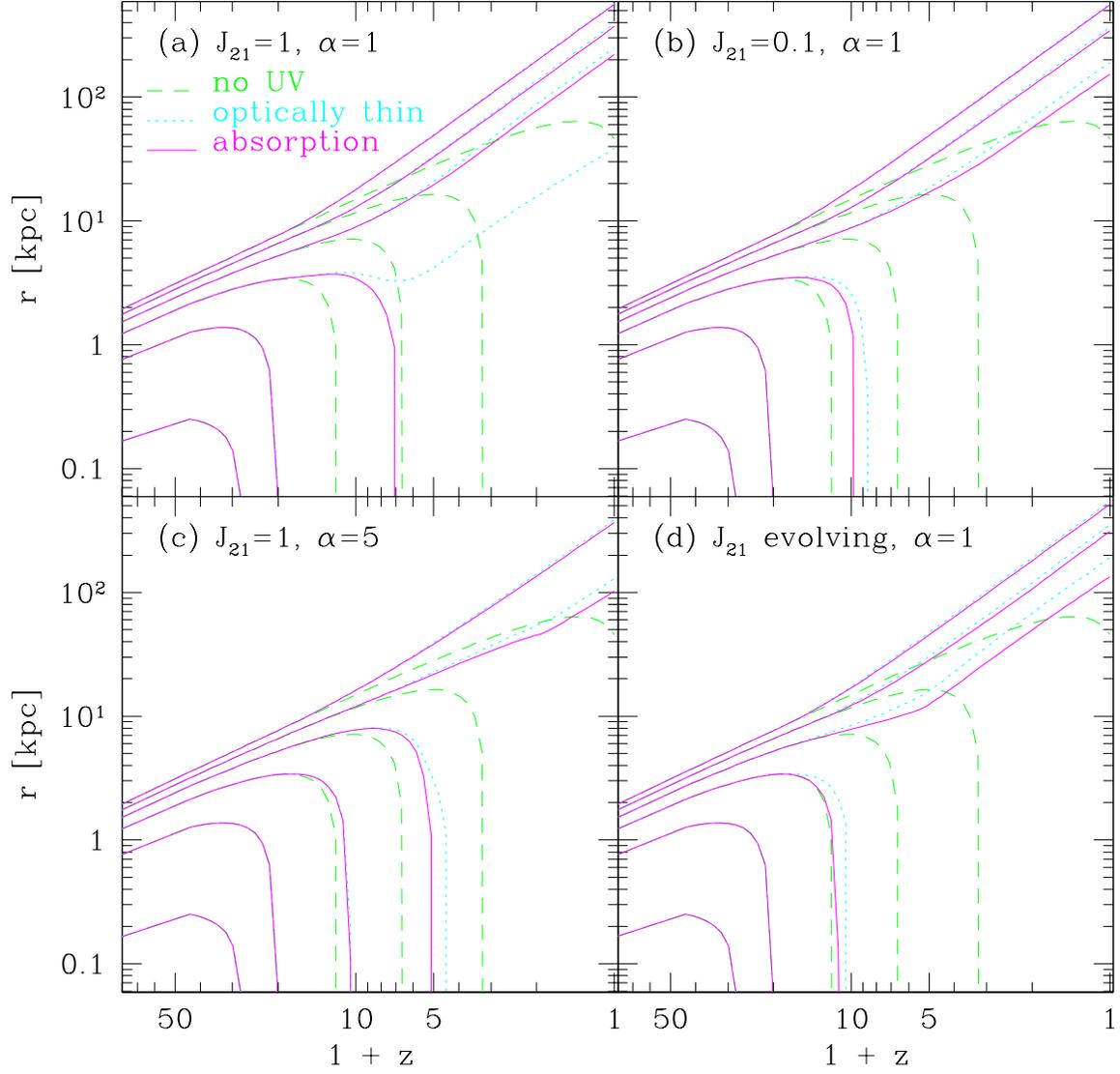}
\caption{Same as Fig.~\protect\ref{fig:r_z1}\protect, except for the
  high-redshift collapse ($z_{c}= 10$) with $V_c=32$ km s$^{-1}$
  ($M_{\rm cloud} \simeq 5 \times 10^7 \msun$).
\label{fig:r_z3}}
\end{figure}

The effects of the UV background radiation is further weakened if one
goes to even higher redshifts. Figure \ref{fig:r_z3} shows the gas
shell trajectories in the high-redshift collapse ($z_{c} = 10$).  At
$z_{\rm UV}$, about 10\% of $M_{\rm bound}$ have already collapsed,
and a number of shells either have turned around or are about to turn
around. In addition, Compton cooling further promotes the gas collapse
at $z \simgt 7$.

\subsection{Critical mass-scales for the collapse}
\label{sec:crit}

\begin{figure}
\plotone{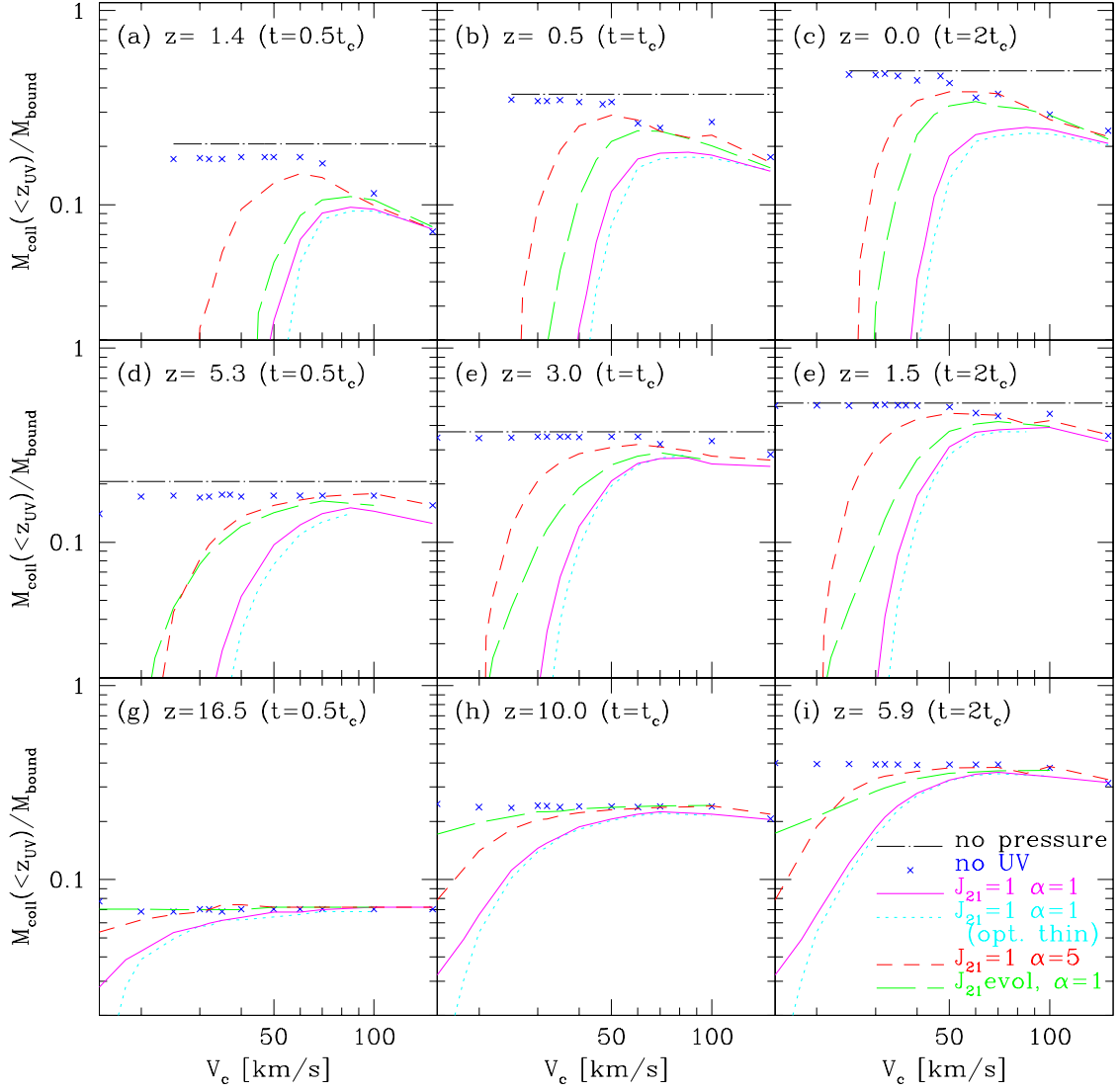}
\caption{The fraction of gas mass collapsed after the onset of the UV
  background $z_{\rm UV}=20$ as a function of circular velocity. Each
  horizontal row traces the time evolution at $t=0.5 t_c$, $t_c$, $2
  t_c$ of runs with $z_c =0.5$ (top panels), $3$ (middle), and $10$
  (bottom). Lines and symbols correspond to different parameters of
  the UV background as shown in the figure (unless indicated,
  absorption is considered).
\label{fig:fc_v}}
\end{figure}

In order to quantify the results presented in \S \ref{sec:dyn}, we
plot in Figure \ref{fig:fc_v} the fraction of gas mass which has
cooled and collapsed to the center after the UV onset $z_{\rm UV}=20$
as a function of circular velocity. The output epochs correspond to
$t=0.5t_c$, $t_c$, $2t_c$, where $t_c = t_0/(1+z_c)^{3/2}$ and $t_0$
is the present age of the universe. In the low and middle redshift
collapses ($z_c=0.5$ and 3), there exists a sharp cut-off in the
collapsed fraction and this threshold lies in the range $V_c = 20 \sim
50$ km s$^{-1}$ depending on different assumptions on the UV background. In
the high-redshift collapse ($z_c=10$), on the contrary, the threshold
is not so distinct and smaller cloud is able to collapse against the
external UV.

\begin{figure}
\plotone{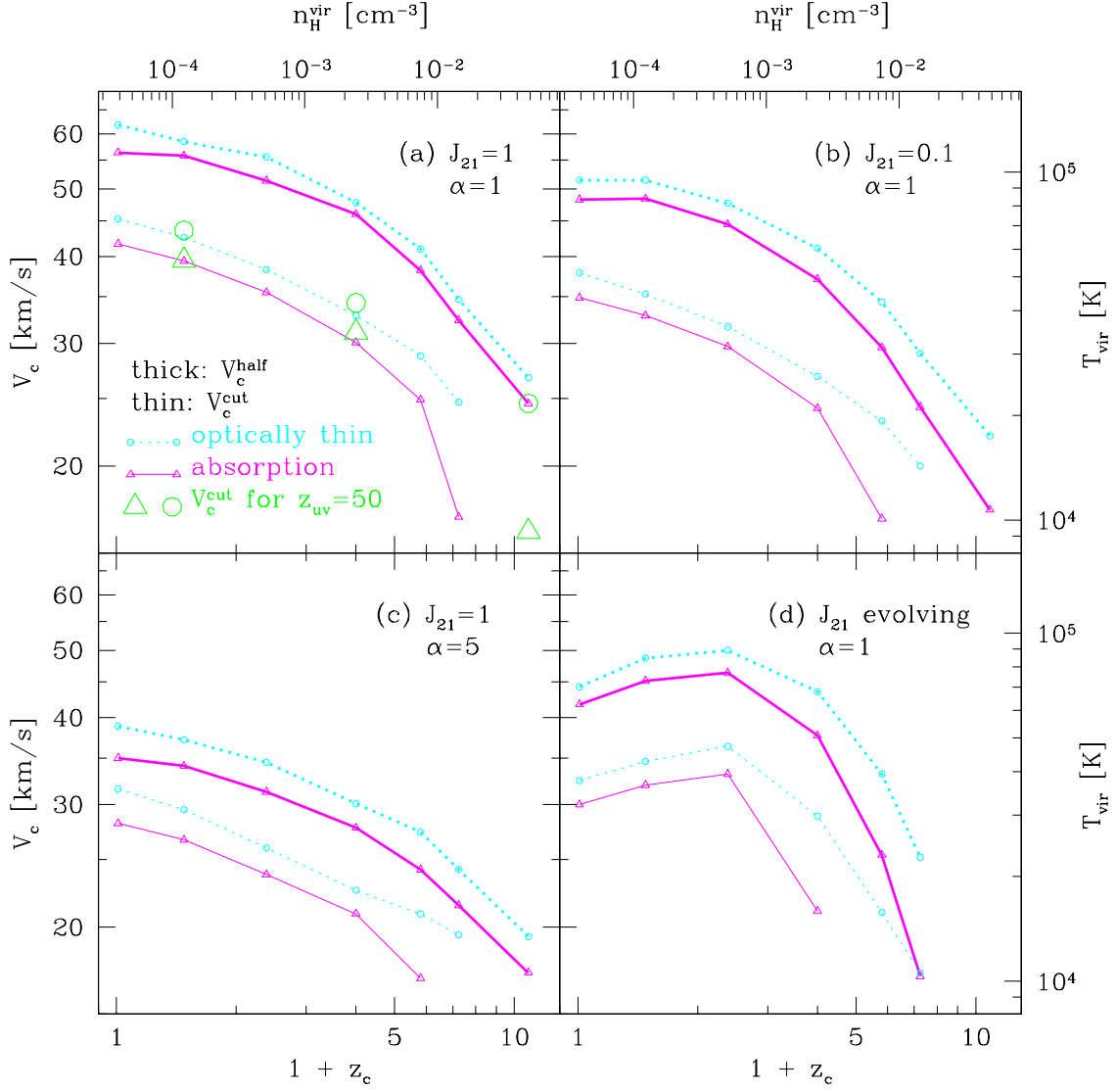}
\caption{The critical circular velocity $V_c^{\rm cut}$ (thin lines)
  and $V_c^{\rm half}$ (thick lines) of the collapse (see text for
  definitions) with absorption (triangles, solid lines) and in the
  optically thin case (circles, dotted lines).  (a) $J_{21}=1$,
  $\alpha=1$, (b) $J_{21}=0.1$, $\alpha=1$, (c) $J_{21}=1$,
  $\alpha=5$, and (d) evolving $J_{21}$, $\alpha=1$.  Larger symbols
  in panel (a) indicate $V_c^{\rm cut}$ in the case of $z_{\rm
  UV}=50$.
\label{fig:v_z}}
\end{figure}

We define the cut-off velocity $V_c^{\rm cut}$ and the half-reduction
velocity $V_c^{\rm half}$, respectively, as the velocities at which
the mass collapsed between $z_{\rm UV}$ and $z_c$ is 0.05 and 0.5 of
that with no UV. Figure \ref{fig:v_z} illustrates that these
quantities depend most largely on the UV spectral index $\alpha$, the
evolution of the UV flux, and the collapsing redshift $z_c$ (we have
added to this figure the results for the runs with $z_c =0$, $1.4$,
$4.8$ and $6.2$). Compared to the optically thin cases, the adoption
of the radiative transfer systematically lowers $V_c^{\rm cut}$ and
$V_c^{\rm half}$ by about 5 km s$^{-1}$.

For a constant UV flux (Figs \ref{fig:v_z} a--c), both $V_c^{\rm cut}$
and $V_c^{\rm half}$ decrease with increasing $z_c$, because of higher
gas density and stronger Compton cooling.  At $z_c\simgt 7$, the
central part of the cloud begins to collapse prior to $z_{\rm UV}$ and
$V_c^{\rm cut}$ falls below 15 km s$^{-1}$, corresponding to $T_{\rm
vir}\sim 8000$K (note that $V_c^{\rm cut}$ and $V_c^{\rm half}$ are
evaluated from the amount of gas cooled after the UV onset). We simply
omit the data points from the figures at such low velocities, since it
is beyond the scope of our current framework to study these regimes
where the molecular cooling dominates.

In the presence of the UV evolution (Fig. \ref{fig:v_z}d), $V_c^{\rm
cut}$ and $V_c^{\rm half}$ rise rapidly with time at high redshifts
and then begin to drop at low redshifts. However, the peaks of these
quantities are at $z_c \sim 1.5$, which is much later than the peak of
the UV flux ($z = 3-6$). This is because the kinetic energy of
expanding gas particles delays the cloud collapse significantly even
after the UV flux starts to decline. The late infall of gas after the
decline of the UV background is therefore suppressed until $z_c \simlt
1$.

So far we have fixed the onset of the UV background at $z_{\rm UV}
=20$. This choice is rather arbitrary given the large uncertainties in
the reionization history of the universe.  A possibility of much later
reionization (smaller $z_{\rm UV}$) can be covered effectively by the
UV evolution of the form (\ref{eq:uvevol}).  In order to examine an
alternative possibility of much earlier reionization, we performed a
few runs with $z_{\rm UV}=50$ and added the resultant $V_c^{\rm cut}$
to Figure \ref{fig:v_z}(a). The differences from the fiducial $z_{\rm
UV}=20$ runs are significant only for the high-redshift collapse ($z_c
= 10$), because the gas shells infalling at $z = 20 - 50$ are mainly
affected.  At such high redshifts, the effect of the radiative 
transfer of UV photons becomes large because the cloud density is very
high. 

\begin{figure}
\plotone{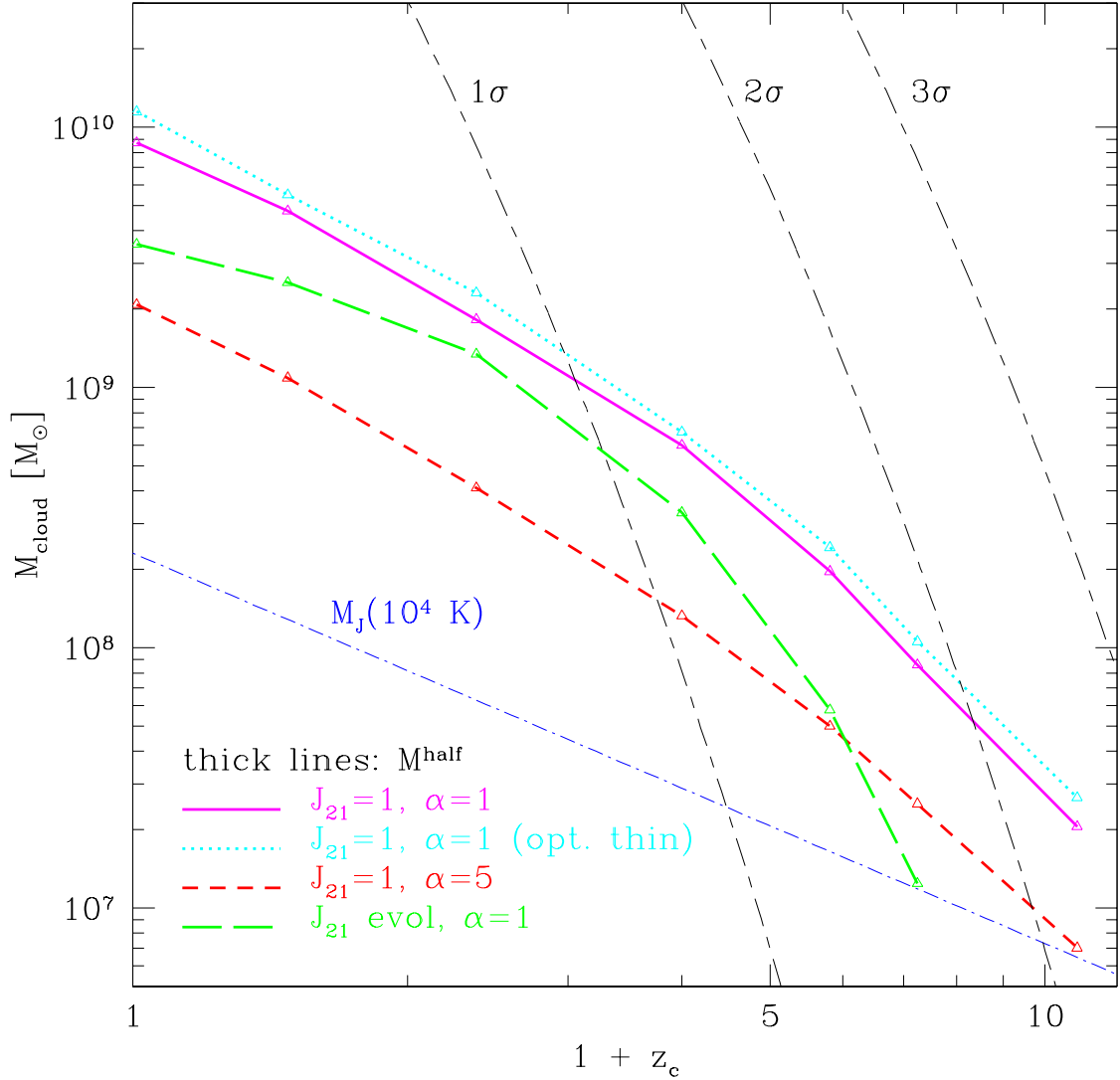}
\caption{The baryon mass corresponding to $V_c^{\rm half}$ as a
  function of collapse epoch $z_c$ (thick lines). Different parameters
  of the UV background are assumed as shown in the figure (unless
  indicated, absorption is considered). Also plotted are the Jeans
  mass with $T=10^4$ K (thin dot-short-dashed lines) and the masses
  corresponding to 1,2,3$\sigma$ density perturbations (thin
  dot-long-dashed lines) in the standard CDM universe with
  $\Omega_0=1$, $h=0.5$, $\Omega_b=0.1$, and $\sigma_8=0.6$.
\label{fig:mc_z}}
\end{figure}

Figure \ref{fig:mc_z} shows the baryon mass $M^{\rm half}$
corresponding to $V_c^{\rm half}$ as a function of $z_c$. Also plotted
for reference are the Jeans mass with $T=10^4$K and the masses
corresponding to 1,2,3$\sigma$ density perturbations in the standard
CDM universe with $\Omega_0=1$, $h=0.5$, $\Omega_b=0.1$, and
$\sigma_8=0.6$.  The cloud well above $T_{\rm vir}=10^4$K is
significantly affected by the UV background radiation. For a constant
UV flux, $M_{\rm b}^{\rm half}$ rises at a rate faster than
$(1+z_c)^{-3/2}$, which is expected for the Jeans mass with constant
temperature. This is because $V_c^{\rm half}$ increases with
decreasing $z_c$ as shown in Figures \ref{fig:v_z} (a)--(c).  For the
evolving UV, $M_{\rm b}^{\rm half}$ keeps growing even after the
decline of the UV background and the growth slows down considerably
only at $z\simlt 1$.

Figure \ref{fig:mc_z} further indicates that the collapse of $\simlt 2
\sigma$ density fluctuations in the standard CDM universe is highly
suppressed at low masses by the UV background radiation.  For example,
the cooled mass from a $1\sigma$ fluctuation is reduced to less than
50\% at $M_{\rm cloud} \simlt 10^9 \msun$ and $z_c \simgt 2$ for
$(J_{21},\alpha)=(1,1)$, and at $M_{\rm cloud} \simlt 2 \times 10^8
\msun$ and $z_c \simgt 3$ for $(J_{21},\alpha)=(1,5)$ (see also \S
\ref{sec:cosm} for more discussion on the implications on structure
formation).

\section{Discussion}

\subsection{Comparison with analytical estimates}

\begin{figure}
\plotone{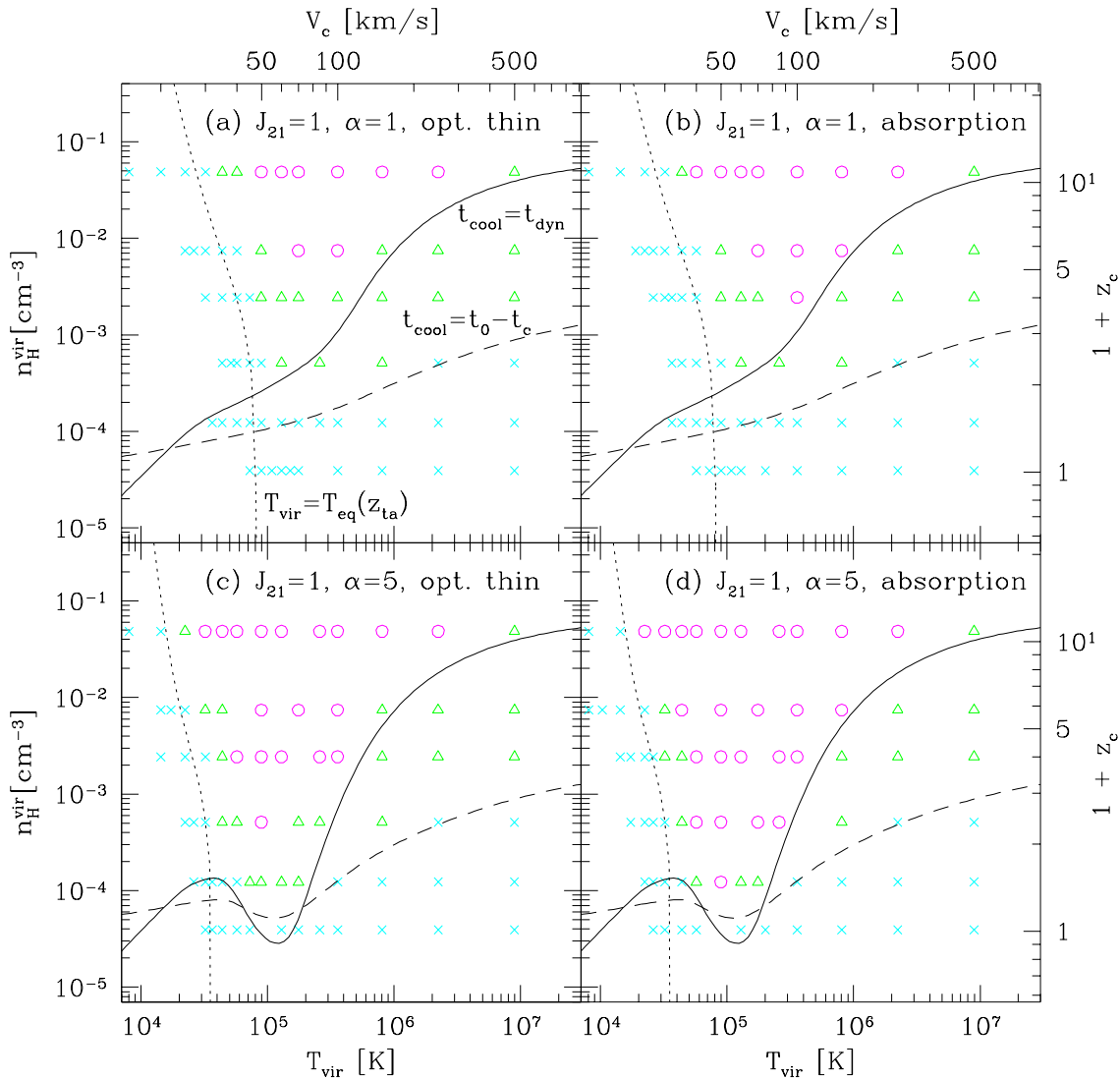}
\caption{Density--temperature diagrams from our simulations and
analytical estimates. Symbols indicate the virial density and
temperature of simulated clouds in which the cooled gas mass reaches
$M_{\rm cloud}$ before $t_c + t_{\rm dyn}$ (circles), before the
present age of the universe $t_0$ (triangles), and otherwise
(crosses); (a) $J_{21}=1$, $\alpha=1$, optically thin case, (b)
$J_{21}=1$, $\alpha=1$, with absorption, (c) $J_{21}=1$, $\alpha=5$,
optically thin case, and (d) $J_{21}=1$, $\alpha=5$, with absorption.
Lines show the analytic relations $t_{\rm cool}=t_{\rm dyn}$ evaluated
at $z_c$ (solid), $t_{\rm cool}=t_0-t_c$ at $z_c$ (dashed), and
$T_{\rm vir}=T_{\rm eq}$ at $z_{\rm ta}$, with the same UV parameters
as the simulations except for assuming that the medium is optically
thin in all cases.
\label{fig:nt}}
\end{figure}

The results of our simulations are compared with analytical estimates
on the density--temperature diagrams (e.g. Efstathiou 1992; Chiba \&
Nath 1994) in Figure \ref{fig:nt}.  We divide the simulation results
into three categories and denote their virial density $n_{\rm H}^{\rm
vir}=3.80\times 10^{-5} (\Omega_b h^2/0.025)(X/0.76) (1+z_c)^3$
cm$^{-3}$ and the virial temperature $T_{\rm vir}$ by different
symbols; 1) {\it efficient cooling} (circles), if the gas mass $M_{\rm
cloud}$ can cool before $t_c + t_{\rm dyn}$, 2) {\it moderate cooling}
(triangles), if $M_{\rm cloud}$ can cool only within the present age
of the universe $t_0$, and 3) {\it inefficient cooling} (crosses), if
otherwise. For comparison, three analytical relations are computed,
using the same UV parameters as the simulations except for assuming
optically thin in all cases; $t_{\rm cool}=t_{\rm dyn}$ (solid line),
$t_{\rm cool}=t_0-t_c$ (dashed), and $T_{\rm vir}=T_{\rm eq}$
(dotted), where $T_{\rm eq}$ is the temperature at which the cooling
and heating rates balance (${\cal H} = {\cal L}$). The former two
analytical relations are evaluated using $n_{\rm H}^{\rm vir}$ at
$z_c$, while the last one using $n_{\rm H}^{\rm ta}= n_{\rm H}^{\rm
vir} / 8$ at the turn-around epoch $z_{\rm ta}$.

Figure \ref{fig:nt} shows that our simulation results under a constant
UV are in reasonable agreement with the analytic estimates defined
above; 1) {\it efficient cooling} lies in the region where $t_{\rm
cool} < t_{\rm dyn}$, $t_{\rm cool} < t_0-t_c$ and $T_{\rm vir} >
T_{\rm eq}$, 2) {\it moderate cooling} mainly lies where $t_{\rm dyn}
< t_{\rm cool} < t_0-t_c$ and $T_{\rm vir} > T_{\rm eq}$, 3) {\it
inefficient cooling} lies where $t_{\rm cool} > t_0-t_c$ or $T_{\rm
vir} < T_{\rm eq}$. Such correspondences degrade near the boundaries
of each region. Compared to the optically thin runs, cooling
efficiency is systematically enhanced to some extent if the absorption
is taken into account.

It should be noted that, in order to achieve above agreements, it is
essential to evaluate the relation $T_{\rm vir}=T_{\rm eq}$ at $z_{\rm
ta}$, not at $z_c$ (if evaluated at $z_c$, the dotted lines in
Fig.~\ref{fig:nt} are shifted downwards by a factor of 8). This is
because the cloud evolution largely depends on the photoionization
prior to the collapse. Roughly speaking, whether the gas can contract
or keeps expanding is determined by the balance between the
gravitational force and the thermal pressure gradient when it is
``maximally exposed'' to the external UV flux, i.e., when the gas
attains the maximum value of $J_{21}/n_{\rm H}$.  For the gas exposed
to the constant UV flux from the linear regime, this corresponds to
the turn around. The above statement also applies in the presence of
the evolution in the UV intensity.

\subsection{Cosmological implications}
\label{sec:cosm}

The suppression of low-mass objects by the UV background radiation has
profound implications on cosmology and galaxy formation. Given the
fraction of cooled gas in objects of different masses and collapse
redshifts from the simulations, we can deduce the abundance
distribution of collapsing objects in the entire
Universe. Specifically, we compute the baryon mass density that cools
and collapses per unit redshift as a function of the collapsed mass
$M_{\rm coll}$ of baryons in a halo and the collapse epoch $z_c$ as
\begin{equation}
  \frac{d^2\rho_{\rm coll}(M_{\rm coll},z_c)}{dM dz} = M_{\rm
      coll}\frac{d^2 N_{\rm halo}^{\rm f}(M_{\rm coll}\Omega_0/f_{\rm
        coll}\Omega_b,z_c)}{dM dz}, 
\end{equation}
where $f_{\rm coll}$ is the collapse fraction of a cloud relative to
the pressure-less case, and $d^2 N_{\rm halo}^{\rm f}(M,z)/dM dz$ is
the comoving number density of halos with mass $M$ that collapse and
form per unit redshift at $z$. We use the latter quantity rather than
a conventional mass function, e.g. that of Press \& Schechter (1974),
because $f_{\rm coll}$ obtained in our simulations is more closely
related to clouds just collapsing at a given redshift. Unfortunately,
there is not yet a fully self-consistent analytical formalism to
compute this quantity, while several approaches have been proposed
(e.g. Bond et al. 1991; Lacey \& Cole 1993; Sasaki 1994; Kitayama \&
Suto 1996a,b; Manrique \& Salvador-Sole 1996). In this paper, as a
working hypothesis, we adopt the halo formation rate given by equation
(15) of Kitayama \& Suto (1996a) with the threshold mass $M_{\rm
f}=M/2$ (see their paper for detail).  We have checked that our
results are qualitatively unchanged by adopting an alternative
approach by Sasaki (1994).

\begin{figure}
\plotone{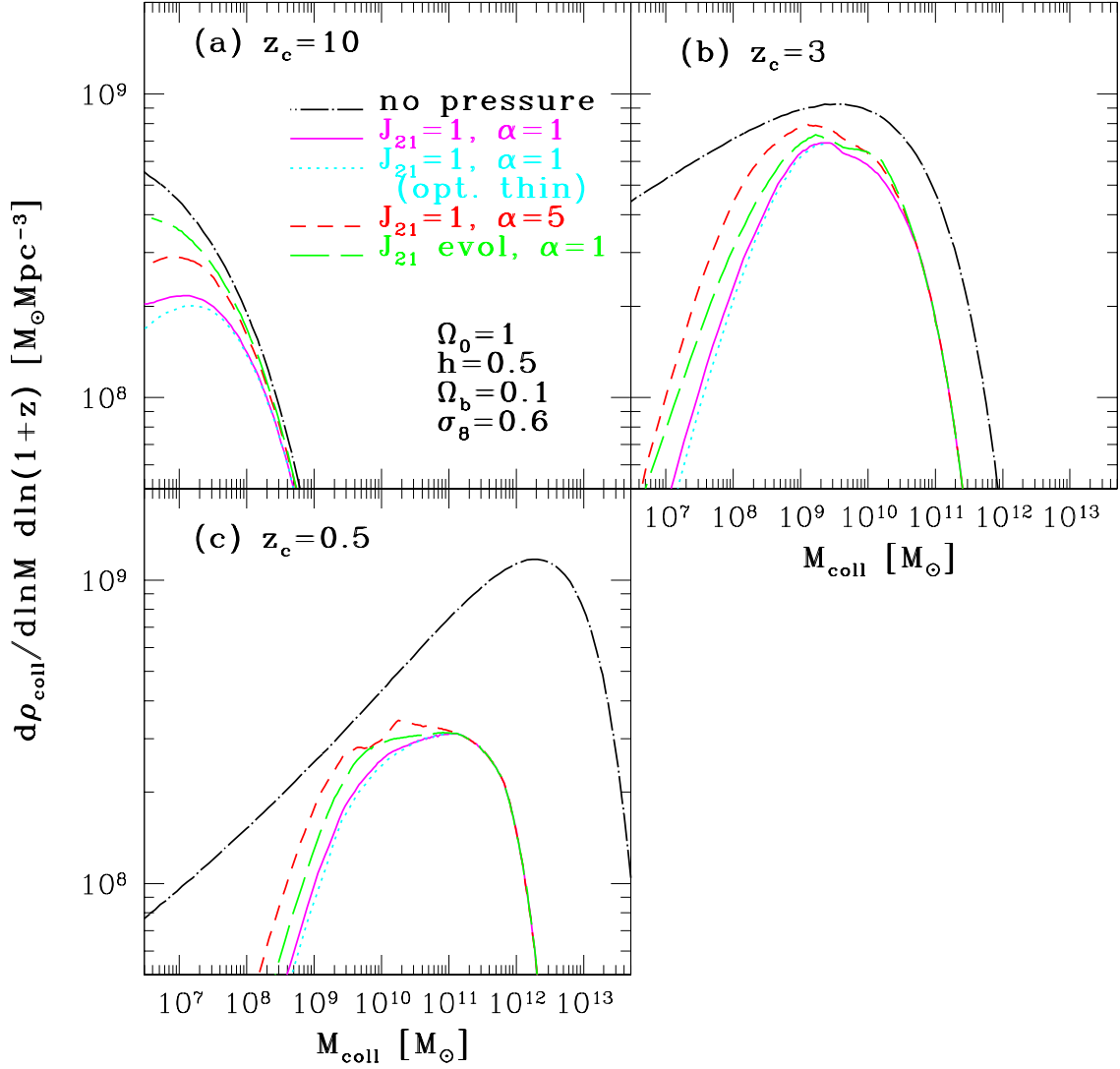}
\caption{Mass distribution of the baryon density that collapses per
  Hubble time at (a) $z_c=10$, (b) $z_c=3$, and (c) $z_c=0.5$ in the
  standard CDM model with $\Omega_0=1$, $h=0.5$, $\Omega_b=0.1$, and
  $\sigma_8=0.6$. Lines indicate the cases of different UV parameters
  or of no pressure as shown in the figure (unless indicated,
  absorption is taken into account).
\label{fig:gasmf}}
\end{figure}

Figure \ref{fig:gasmf} illustrates the baryon mass distribution of
collapsing objects in the standard CDM universe. In the absence of
pressure, collapsing objects have a broad mass distribution with an
increasing fraction of high mass objects at lower redshifts. Once the
gas pressure is taken into account, high mass objects are reduced by
the low cooling efficiency, and low mass objects by the UV background. 
As a result, the abundance distribution at $z_c \simlt 3$ agrees well
with the observed mass range of galaxies. Note that the low mass end
of the distribution has a tail, not a sharp cut-off. This is because
the objects just above $V_c^{\rm cut}$ can have a wide range of
$M_{\rm coll}$ for almost identical $M_{\rm cloud}$. Thus the collapse
of dwarf-sized objects of mass $M_{\rm coll}\simlt 10^9 \msun$ is
still possible at $z_c \sim 0.5$, but such objects are expected to
have lower baryon fraction than normal galaxies. This point can be
tested by future observations.

\begin{figure}
\plotone{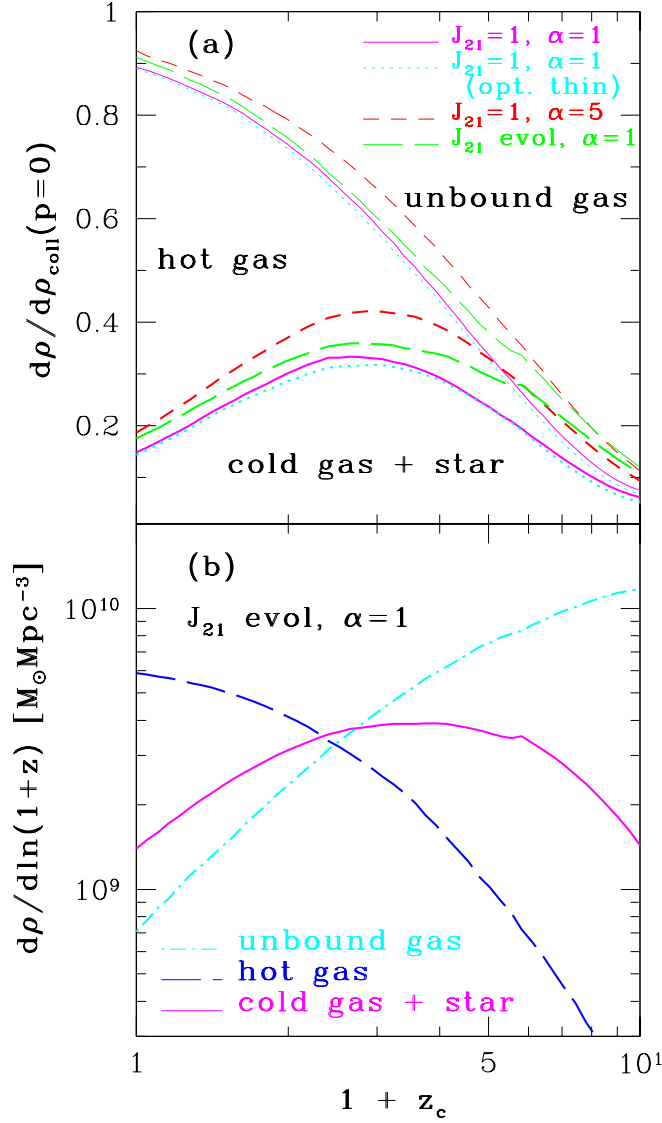}
\caption{(a) The baryon density which cools (thick lines) or becomes
  bounded (thin lines) per Hubble time under the UV background,
  normalized by that which collapses in the absence of gas
  pressure. Lines specify different UV parameters as shown in the
  figure (unless indicated, absorption is considered).  The three
  regions divided by thick and thin lines correspond to ``cold gas +
  star'', ``hot gas'', and ``unbound gas'', respectively.  (b)
  Production rates of ``cold gas + star'', ``hot gas'', and ``unbound
  gas'' for the evolving $J_{21}$ and $\alpha=1$. For both panels, the
  standard CDM model is assumed.
\label{fig:gasmf_z}}
\end{figure}

One can integrate the above distribution over mass to obtain the total
amount of baryon that is collapsing at a given epoch. Figure
\ref{fig:gasmf_z} shows the evolution of this quantity (denoted by
``cold gas + star'').  We have also deduced similarly from our
simulations the baryon mass density that becomes bound by the gravity
of clouds but not yet cooled at $z_c$. Such a component, denoted by
``hot gas'', is mainly contained in objects corresponding to groups
and clusters of galaxies. We further define ``unbound gas'' as the gas
that becomes unbound due to the UV background but would have collapsed
at $z_c$ in the absence of pressure. Figure \ref{fig:gasmf_z}
indicates that the production rate (per Hubble time at a given epoch)
of ``cold gas + star'' has a peak at $z_c \sim 2 - 3$ and it is rather
insensitive to the UV parameters.  This seems to suggest a high
efficiency of star formation activity at these redshifts. On the other
hand, the production rate of ``hot gas'' and that of ``unbound gas''
simply increases and decreases with time, respectively (see also
Barkana \& Loeb 1999; Cen \& Ostriker 1999).

\begin{figure}
\plotone{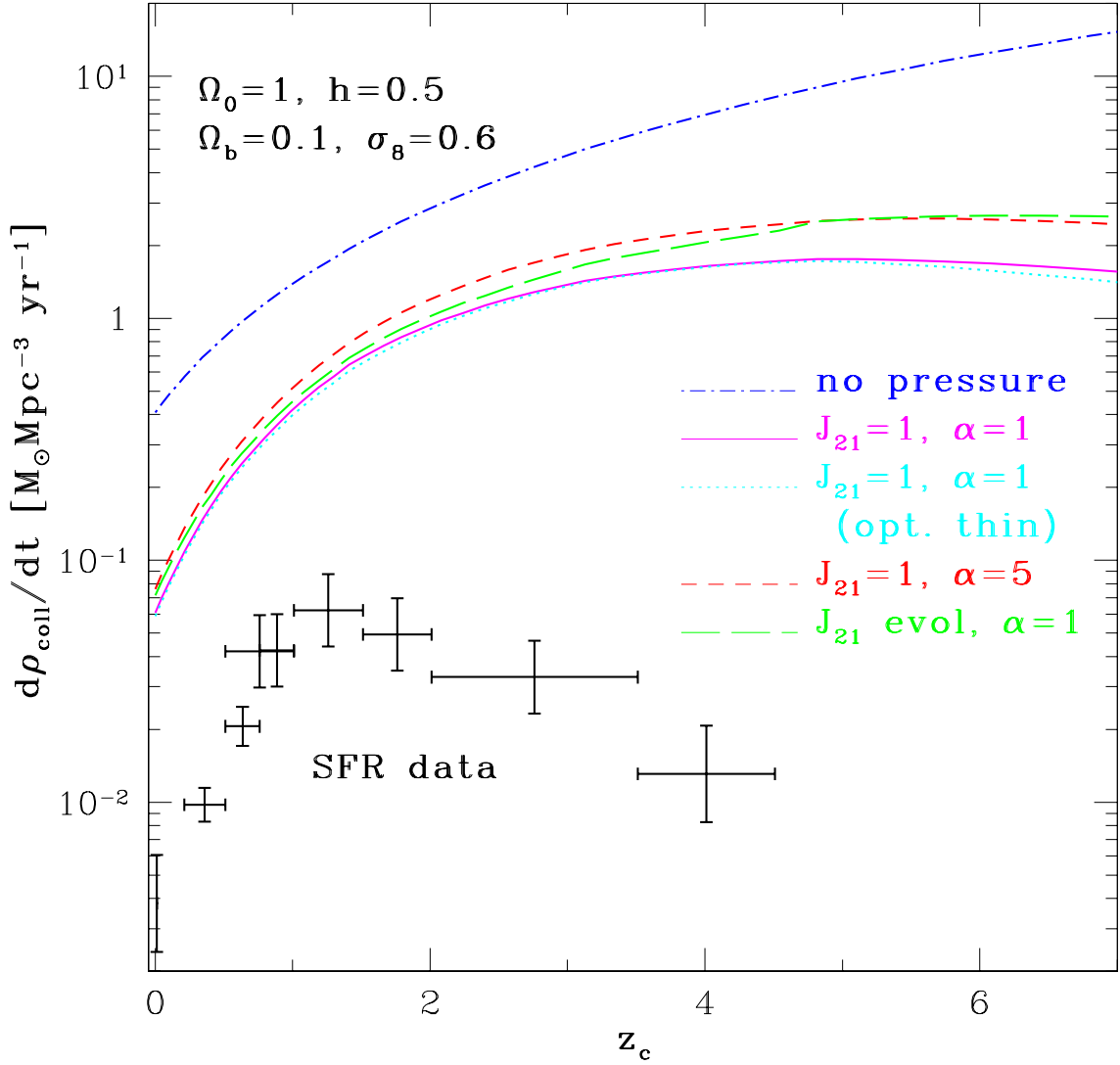}
\caption{The global production rates of ``cold gas + star'' per year
predicted in the standard CDM model (unless indicated, absorption is
considered). Also potted for reference are the observed cosmic star
formation rates (Madau et al. 1996, 1998; Lilly et al. 1996; Connolly
et al. 1997), without corrections for the dust extinction, compiled by
Totani et al. (1997) and Totani (1999, private communication).
\label{fig:sfr_z}}
\end{figure}

Figure \ref{fig:sfr_z} further compares the predicted production rate
(per year) of ``cold gas + star'' in the standard CDM universe with
the observed cosmic star formation rate (SFR, Madau et al. 1996, 1998;
Lilly et al. 1996; Connolly et al. 1997) compiled by Totani et
al. (1997) and Totani (1999, private communication). This comparison
is only demonstrative, since the predicted curves simply set an upper
limit to the global SFR in the universe and their normalization can be
shifted in proportion to the baryon density parameter $\Omega_b$.
Nonetheless, it is noticeable that the predicted production rates of
``cold gas + star'' under the UV background show a steep rise at $z
\simlt 1$, in good qualitative accordance with what is actually seen
in the SFR data. At higher redshifts, they continue to rise up to $z
\sim 4$ and then flatten. This is somewhat different from a rather
sharp decline at $z \sim 2-4$ in the observed SFR. It should be kept
in mind, however, that these SFR data are still very uncertain due to
the dust extinction and may be shifted upward by large factors
(e.g. Pettini et al.  1998). Recent detections of star forming
galaxies in the submm band also suggest rather high SFR of $\simgt 2
\times 10^{-1} h \msun \mbox{yr}^{-1} \mbox{Mpc}^{-3}$ at $z \sim 2-4$
(Smail, Ivison \& Blain 1997; Hughes et al 1998; Berger et al. 1998).
More detailed discussion on this subject should await higher precision
data from future infrared observations.

\subsection{Imprints on quasar absorption lines}

\begin{figure}
\plotone{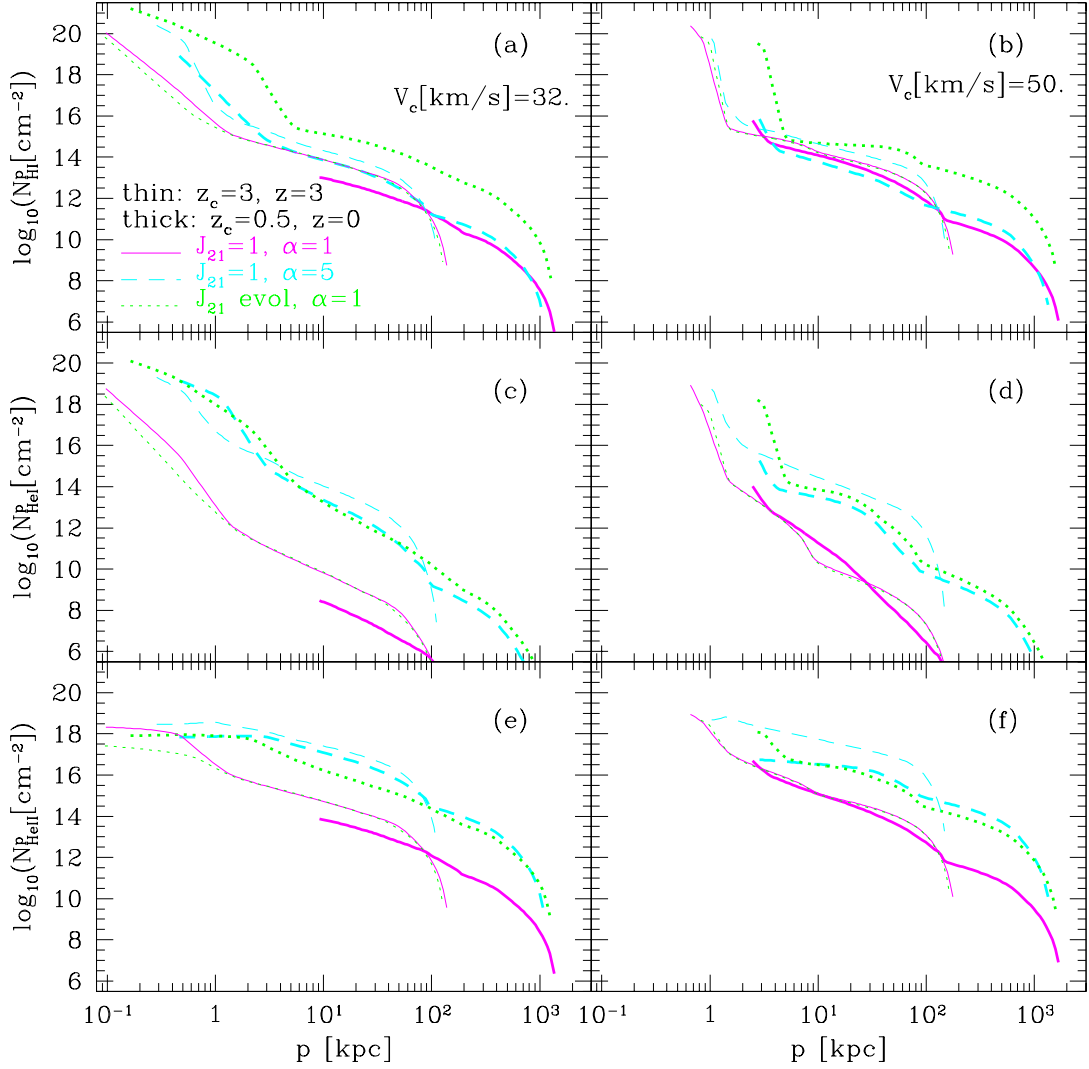}
\caption{Line of sight column densities through the cloud $N^{\rm
 p}_j$ ($j=$HI, HeI, HeII) as a function of impact parameter $p$ for
 $V_c=32$ km s$^{-1}$ (left panels) and $V_c=50$ km s$^{-1}$ (right). Thick lines
 are the clouds with $z_c=3$ viewed at $z=3$, while thin lines those
 with $z_c=0.5$ viewed at $z=0$. Choices of the UV parameters (all
 with absorption) are as indicated in the figure. \label{fig:cl1}}
\end{figure}

The effects of the UV background radiation on the intergalactic medium
can be observed most directly by means of absorption lines of quasar
spectra.  In order to discuss the impacts of current results on such
observations, we plot in Figure \ref{fig:cl1} the line of sight column
densities through the cloud $N^{\rm p}_j$ ($j=$HI, HeI, HeII) as a
function of impact parameter $p$. Here, we restrict the observability
of these elements to the column densities $N^{\rm p}_j >
10^{14}$cm$^{-2}$, and define the critical impact parameter $p_{\rm
crt}$ within which the column density is greater than
$10^{14}$cm$^{-2}$. As the column density of each element is too small
to be detected except at the central region in the case of $J_{21}=1$
and $\alpha=1$, we focus on the other two cases of UV parameters shown
in Figure \ref{fig:cl1}. In Tables \ref{tab:cl1} and \ref{tab:cl2}, we
summarize $p_{\rm crt}$ and the ratio of its square value, which
corresponds to the ratio of area, i.e., the relative expected number.
These tables suggest that a large number of helium forest lines are
expected especially under the soft UV spectrum. Recent observations by
HST in fact indicate the detections of numerous HeII lines (Hogan,
Anderson \& Rugers 1997; Reimers et al.1997; Anderson et al. 1999).

If we fit the lines in Figure \ref{fig:cl1} at $N^{\rm p}_j > 
10^{14}$cm$^{-2}$ by a single power-law $N^{\rm p}_j \propto p^{-n}$,
then the column density distribution of each absorption line is
written as $d{\cal N}/dN^{\rm p}_j \propto ({N^{\rm p}_j})^{-\beta}$
with $\beta=(n+2)/n$. In Table \ref{tab:cl3}, we summarize $n$ and
$\beta$ values.  Since these are crude numbers, we merely present the
values in the case of evolving $J_{21}$ at $z_c=0.5$ and ignore the
dependence on redshift and UV parameters. At $N^{\rm p}_{\rm HI} >
10^{16}$cm$^{-2}$, the column density distribution of HeI is similar
to that of HI while that of HeII is a little steeper. These points can
be checked by future observations.

\begin{figure}
\plotone{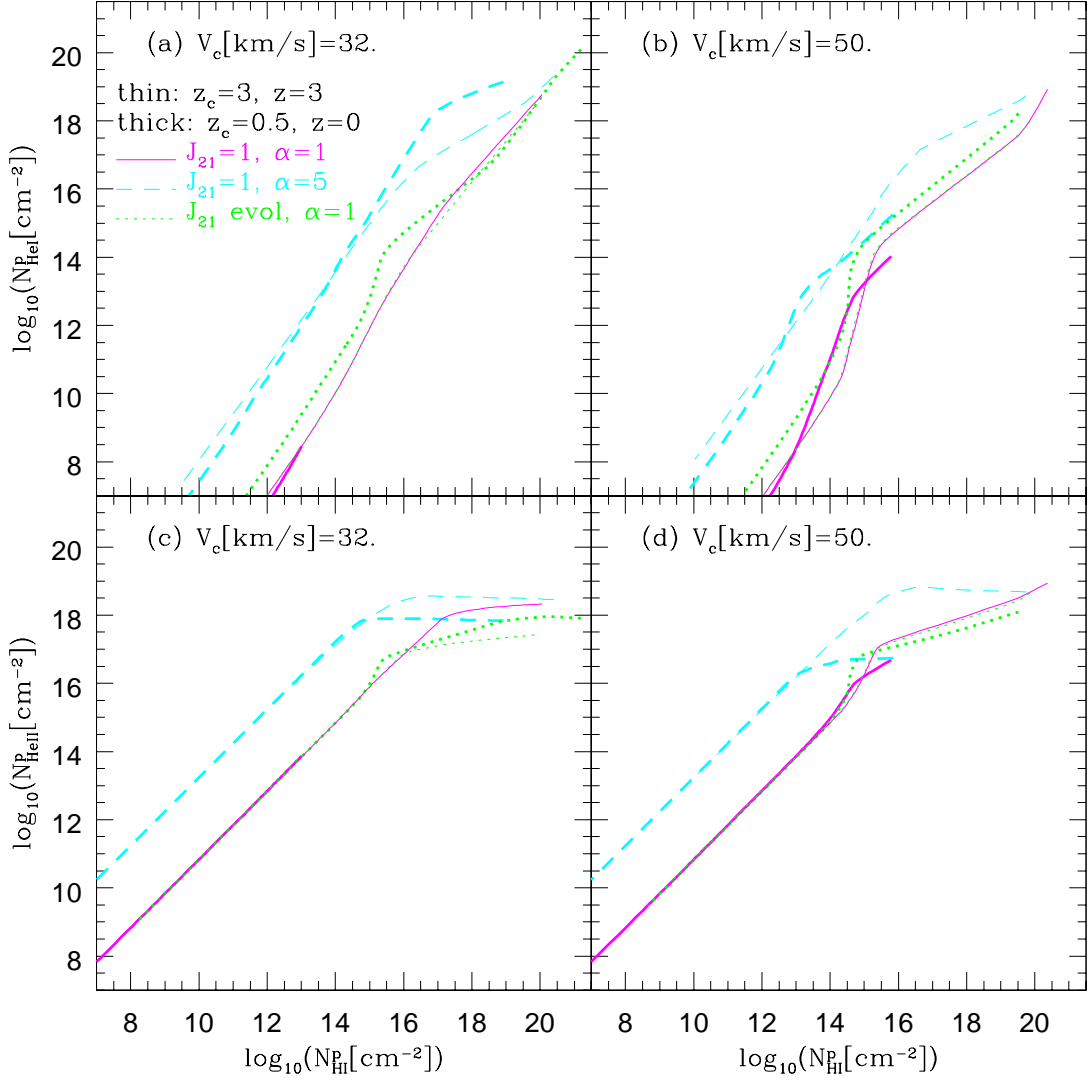}
\caption{Same as Fig.~\protect\ref{fig:cl1}\protect, except that
$N^{\rm p}_{\rm HeI}$ (upper panels) and $N^{\rm p}_{\rm HeII}$
(lower) are plotted against $N^{\rm p}_{\rm HI}$.
\label{fig:cl2}}
\end{figure}

The above results are also clearly seen in Figure \ref{fig:cl2}, which
plots the line of sight column densities of HeI and HeII against the
HI column density. The relations among these column densities are
rather insensitive to $V_c$ and $z_c$, and are summarized in Table
\ref{tab:cl4} (values are given for $V_c=32$ km s$^{-1}$ and $z_c=0.5$). In
order for the HeI lines to be detected at $N^{\rm p}_{\rm HeI} >
10^{14}$cm$^{-2}$, the corresponding HI forest lines are required to
have $N^{\rm p}_{\rm HI} > 10^{14} - 10^{16}$cm$^{-2}$. On the other
hand, HeII forest lines are more easily detectable and are even
saturated at $N^{\rm p}_{\rm HI} > 10^{16}$cm$^{-2}$.

In addition to the high observability of the helium absorption lines
at the UV wavelengths, our results further suggest an interesting
feature in the HI absorption lines at high column densities.  When the
line of sight passes near the central region of a cloud that has just
collapsed, the gas is expected to be neutral but the temperature can
be as high as $10^4$K, because the time-scale of ionization is longer
than that of the photoionization heating (eq.~[\ref{eq:heation}]). It
should be kept in mind that if the hydrogen molecules, ignored in the
present paper, are present, they could allow cooling below $10^4$K. In
any case, the existence of the warm neutral gas could be checked by
the HI absorption lines with a large Doppler parameter and the 21cm
emission line with high spin temperature. Incidentally the recent
observations of damped Ly$\alpha$ systems (Lane et al. 1998; Chengalur
\& Kanekar 1999) suggest the detection of neutral gas with the spin
temperature $ \sim 10^3$K, which is much higher than what is found in
normal spiral galaxies.

\section{Conclusions}

We have shown that the formation of the sub-galactic clouds is greatly
prevented by the UV background radiation even if the transfer of the
external UV photon is taken into account. Within the range of
parameters investigated in this paper ($J_{21} \leq 1$, $\alpha=1,5$,
$z_c=0-10$), the complete suppression of collapse occurs for the
clouds with circular velocities typically in the range $V_c \sim 15 -
40$ km s$^{-1}$ and the 50\% reduction in the cooled gas mass with $V_c \sim
20 - 55$ km s$^{-1}$. These values depend most sensitively on the collapse
redshift $z_c$ and the slope of the UV spectrum $\alpha$.

The evolution of the UV background also affects the above thresholds
in a significant manner. The decline of the UV intensity at $z \simlt
3$ can decrease the threshold circular velocities at lower redshifts.
This effect, however, is delayed until $z \simlt 1.5$ due to the
kinetic energy of gas particles attained at higher redshifts.  In
fact, whether the gas can contract or keeps expanding is roughly
determined by the balance between the gravitational force and the
thermal pressure gradient when the gas attains the maximum value of
$J_{21}/n_{\rm H}$.

As far as the evolution of the gas down to $T \sim 10^4$K is
considered, the radiative transfer of the ionizing photons has the
moderate effect. Compared to the optically thin case, the absorption
of the UV photons by the intervening medium systematically lowers the
above threshold values by $\Delta V_c \sim 5$ km s$^{-1}$.  Once the
evolution below $T \sim 10^4$K is taken into account, incorporating
the formation and destruction of hydrogen molecules, the radiative
transfer is expected to be of greater significance (Haiman, Rees \&
Loeb 1996, 1997). We will investigate the dynamical evolution of the
gas clouds in this regime in future publications (Kitayama et al. in
preparation; see also Susa \& Umemura 1999).

Our calculations are in good accordance with those of Thoul \&
Weinberg (1996), for the the same set of parameters, i.e. in the
optically thin case and $z_c = 2 - 5$. This gives an important
cross-check of the current results and confirms that they are
insensitive to the different choices of initial conditions and central
boundary conditions between their calculations and ours (see \S 2.4
and \S 4.1 of Thoul \& Weinberg 1995).

Based on the results of numerical simulations, we have predicted the
global production rates of cold gas, hot gas, and unbound gas in the
standard CDM universe. The abundance distribution of the cold gas
matches well the observed mass ranges of dwarfs and galaxies. The
global production rate of cooled gas is found to rise steeply from the
local universe to $z \sim 2-3$, indicating a higher efficiency of star
formation activity at high redshifts.

We further predict that a large number of the HeII and HeI forest
lines may arise in the quasar spectra at the UV wavelengths, which can
be detected by HST and future space missions. Such observations should
provide powerful probes of the physical state of the intergalactic
medium, such as the gas kinematics, and the UV background spectrum
(e.g. Sargent et al. 1980; Rauch 1998).  In addition, the existence of
the warm neutral gas is inferred, due to the high efficiency of
heating over ionization. The temperature of such gas can be as high as
$\sim 10^4$K and it may be related to the high spin temperatures
suggested from the 21cm absorption line observations in the nearby
damped Ly$\alpha$ systems (Lane et al. 1998; Chengalur \& Kanekar
1999). In testing these predictions, it is essential to perform the
multi-line analyses of the absorption systems, which will become
possible with greater precisions in near future.

\bigskip 
\bigskip 
\bigskip 

We thank Naoteru Gouda, Izumi Murakami, Tatsushi Suginohara, Hajime
Susa, Yasushi Suto and Masayuki Umemura for discussions, and the
referee, Rennan Barkana, for helpful comments on the manuscript. We
are also grateful to Masahiro Kawasaki and Hideyuki Suzuki for
suggestions on the numerical code, and Tomonori Totani for providing
the SFR data.  T.K. acknowledges a fellowship from Japan Society for
the Promotion of Science.

\bigskip 
\bigskip
\bigskip 

\appendix
\section{An analytical solution for radiative transfer: photoionization
  coefficients and heating rates of primordial gas}

We derive analytical formulae for the photoionization coefficients and
heating rates of a radiation field penetrating into primordial gas
aligned in a plane parallel geometry, with an arbitrary density
profile.  The frequency/direction-dependent radiative transfer due to
the absorption by multiple species is explicitly taken into account.
While the gas composed of atomic hydrogen and helium is considered
here, the formalism can be readily extended to include other
species. The derived formulae are applicable to a variety of problems,
e.g., the reionization of the universe, and the photoionization of a
mini-pancake or low-metallicity gas in galactic haloes.

\subsection{Derivation}

Suppose that an isotropic incident radiation field is processed
through a gas slab composed mainly of atomic hydrogen and helium. For
simplicity, we neglect the emission and scattering of photons by the
intervening medium and only consider the absorption above the
ionization energy of each species. We model the incident radiation
spectrum as a power law with an index $\alpha$:
\begin{equation} 
  I_0(\nu) = I_0(\nu_{\rm HI}) \sbkt{\frac{\nu}{\nu_{\rm
        HI}}}^{-\alpha},
\end{equation} 
where $I_0(\nu_{\rm HI})$ is the specific intensity at the ionization
frequency of neutral hydrogen $\nu_{\rm HI}$.  Then the processed mean
intensity at an arbitrary point inside the slab is written as
\begin{equation} 
  J(\nu) = \frac{1}{4\pi} \int d \omega I_0(\nu_{\rm HI})
  \sbkt{\frac{\nu}{\nu_{\rm HI}}}^{-\alpha} e^{-\tau(\nu,\omega)}, 
\label{eq:j}
\end{equation} 
using the optical depth $\tau(\nu,\omega)$ along a photon path $s$
from the slab boundary to the point in the incident direction
$\omega$:
\begin{equation} 
  \tau(\nu,\omega) = \sum_i \sigma_i(\nu) \int_\omega n_i ds, 
\label{eq:tau}
\end{equation} 
where $\sigma_i(\nu)$ and $n_i$ are the photoionization cross section
and the number density of the species $i$ (= 1,2,3 or HI, HeI, HeII,
in the ascending order in its ionization energy), respectively.  For
simplicity, we approximate the cross sections by a single power-law:
\begin{equation} 
  \sigma_i(\nu) = \sigma_i(\nu_i) \sbkt{\frac{\nu}{\nu_i}}^{- \eta_i}
  \Theta(\nu-\nu_i),
\label{eq:cross}
\end{equation}  
where $\Theta(x)$ is the Heviside step function, and the amplitude
$\sigma_i(\nu_i)$, the index $\eta_i$, and the ionization energy $h
\nu_i$ are taken from Osterbrock (1989) and listed in Table
\ref{tab:cross}. Note that the following formalism can be readily
extended and applied as long as cross sections are expressed as a
superposition of power-laws.

Using the above expressions, the photoionization coefficients and
heating rates for the species $j$ are written as 
\begin{eqnarray} 
\label{eq:coeff}
  \Gamma_j &=& \int_{\nu_{j}}^\infty \frac{4 \pi J(\nu)}{h \nu}
  \sigma_j(\nu) d\nu, \nonumber \\
&=& \frac{\sigma_j(\nu_j) I_0(\nu_j)}{h} \int d\omega
\int_{\nu_j}^\infty \frac{d\nu}{\nu}
\sbkt{\frac{\nu}{\nu_j}}^{-\alpha-\eta_j} e^{-\tau(\nu,\omega)}, \\
 {\cal H}_j &=& \int_{\nu_{j}}^\infty \frac{4 \pi J(\nu)}{h \nu}
  \sigma_j(\nu) (h\nu - h\nu_j)d\nu \nonumber \\
&=& h \nu_j \lbkt{\frac{\sigma_j(\nu_j) I_0(\nu_j)}{h} \int
  d\omega \int_{\nu_j}^\infty \frac{d\nu}{\nu_j}
  \sbkt{\frac{\nu}{\nu_j}}^{-\alpha-\eta_j} e^{-\tau(\nu,\omega)} -
  \Gamma_j}. 
\label{eq:rate}
\end{eqnarray} 
Note that ${\cal H}_j$ defined above is related to the heating rate
per unit volume used in the main text (eq. [\ref{eq:energy}]) by
${\cal H} = \sum n_j {\cal H}_j$.  In general, $\tau(\nu,\omega)$ is a
complicated function of $\nu$ as it is the sum of components with
different spectral indices (eqs~[\ref{eq:tau}],[\ref{eq:cross}]).
However, the frequency integrations in equations (\ref{eq:coeff}) and
(\ref{eq:rate}) are separated into intervals as
\begin{equation}
\int_{\nu_j}^\infty d\nu ~~~ \longrightarrow ~~~ \sum_{i=j}^3
\int_{\nu_i}^{\nu_{i+1}} d\nu  ~~\mbox{ with $\nu_4 \equiv\infty$}, 
\end{equation} 
and it will not be a bad approximation to adopt for each interval
\begin{equation}
  \tau(\nu,\omega) \simeq \tau(\nu_i,\omega)
 \sbkt{\frac{\nu}{\nu_i}}^{-\eta_i^{\rm eff}} 
~~~~\mbox{at $\nu_i \leq \nu < \nu_{i+1}$,}  
\label{eq:tauapp}
\end{equation} 
where $\tau(\nu_i,\omega)$ takes the sum over all species at $\nu_i$
(eq.[\ref{eq:tau}]), while the effective index $\eta_i^{\rm eff}$ is
set equal to the index $\eta$ of the species which makes the dominant
contribution to $\tau(\nu_i,\omega)$. Then the frequency integrations
in equations (\ref{eq:coeff}) and (\ref{eq:rate}) are performed
analytically to give (see also Tajiri \& Umemura 1998; Susa \& Umemura
1999)
\begin{eqnarray} 
\label{eq:coeff3}
  \Gamma_j &=& \frac{\sigma_j(\nu_j) I_0(\nu_j)}{h} 
\int d\omega  \sum_{i=j}^3 \frac{1}{\eta_i^{\rm eff}} 
\left\{\frac{\gamma\!\sbkt{\beta_{ji}, \tau(\nu_i,\omega)}}
{\tau^{\beta_{ji}}(\nu_i,\omega)}
\sbkt{\frac{\nu_i}{\nu_j}}^{-\beta_{ji}\eta_i^{\rm eff}} \right. \nonumber \\
& & \left. - \frac{\gamma\!\sbkt{\beta_{ji}, \tilde{\tau}(\nu_{i+1},\omega)}}
{\tilde{\tau}^{\beta_{ji}}(\nu_{i+1},\omega)}
\sbkt{\frac{\nu_{i+1}}{\nu_j}}^{-\beta_{ji}\eta_i^{\rm eff}}\right\}, \\
{\cal H}_j &=& h \nu_j \left[\frac{\sigma_j(\nu_j) I_0(\nu_j)}{h} \int d\omega
\sum_{i=j}^3 \frac{1}{\eta_i^{\rm eff}} 
\left\{\frac{\gamma\!\sbkt{\beta_{ji}^\prime, \tau(\nu_i,\omega)}}
{\tau^{\beta_{ji}^\prime}(\nu_i,\omega)}
\sbkt{\frac{\nu_i}{\nu_j}}^{-\beta_{ji}^\prime\eta_i^{\rm eff}} \right.\right.
\nonumber \\
& & \left.\left. -\frac{\gamma\!\sbkt{\beta_{ji}^\prime, 
\tilde{\tau}(\nu_{i+1},\omega)}}
{\tilde{\tau}^{\beta_{ji}^\prime}(\nu_{i+1},\omega)}
\sbkt{\frac{\nu_{i+1}}{\nu_j}}^{-\beta_{ji}^\prime\eta_i^{\rm eff}}\right\}
- \Gamma_j\right],  
\label{eq:rate3}
\end{eqnarray} 
with
\begin{equation}
\beta_{ji}\equiv \frac{\alpha+\eta_j}{\eta_i^{\rm eff}},~~~~~
\beta_{ji}^\prime\equiv \frac{\alpha+\eta_j-1}{\eta_i^{\rm eff}}, 
\end{equation} 
where $\tilde{\tau}(\nu_{i+1},\omega)$ is an approximated optical
depth computed by equation (\ref{eq:tauapp}) at its upper limit
$\nu_{i+1}$, and  $\gamma(a,x)$ is the incomplete gamma function: 
\begin{equation} 
\gamma(a,x) \equiv  \int_0^x e^{-t} t^{a-1} dt,  ~~~ a>0 . 
\label{eq:gam}
\end{equation} 
In most astrophysically interesting cases, $\beta>0$ and
$\beta^\prime>0$ are both satisfied.

For a sufficiently long slab in a plane parallel symmetry with an
arbitrary density profile, the optical depth given in equation
(\ref{eq:tau}) varies with a direction cosine $\mu = \cos \theta$
($\theta=0$ if perpendicular to the plane) as
\begin{eqnarray} 
  \tau(\nu,\omega(\nu)) & =& \frac{1}{\mu} 
\sum_i \sigma_i(\nu) N_i, \nonumber \\
&=& \frac{1}{\mu} \tau_\bot(\nu)
\label{eq:tau2}
\end{eqnarray} 
where $N_i = \int_{\theta=0} n_i ds$ is the column density
perpendicular to the plane.  Substituting equation (\ref{eq:tau2})
into equations (\ref{eq:j}), (\ref{eq:coeff3}), (\ref{eq:rate3}), and
taking $\int d \omega \rightarrow 2 \pi \int_0^1 d\mu$, we obtain the
following analytical formulae for the photons propagating from {\it
one of the boundaries}:
\begin{eqnarray} 
\label{eq:j2}
  J(\nu) &=& \frac{1}{2} I_0(\nu_{\rm HI})
  \sbkt{\frac{\nu}{\nu_{\rm HI}}}^{-\alpha} 
\lbkt{e^{-\tau_\bot(\nu)} - \tau_\bot(\nu) E_1(\tau_\bot(\nu))}, \\ 
\label{eq:coeff4}
  \Gamma_j &=& \frac{2\pi\sigma_j(\nu_j) I_0(\nu_j)}{h} 
  \sum_{i=j}^3 \frac{1}{\eta_i^{\rm eff}} 
\left\{ f(\beta_{ji},\tau_\bot(\nu_i))
\sbkt{\frac{\nu_i}{\nu_j}}^{-\beta_{ji}\eta_i^{\rm eff}}
\right. \nonumber \\
&& \left.- f(\beta_{ji},\tilde{\tau}_\bot(\nu_{i+1}))
\sbkt{\frac{\nu_{i+1}}{\nu_j}}^{-\beta_{ji}\eta_i^{\rm eff}}\right\}, \\
{\cal H}_j &=& h \nu_j \left[\frac{2\pi\sigma_j(\nu_j) I_0(\nu_j)}{h} 
\sum_{i=j}^3 \frac{1}{\eta_i^{\rm eff}} 
\left\{f(\beta_{ji}^\prime,\tau_\bot(\nu_i))
\sbkt{\frac{\nu_i}{\nu_j}}^{-\beta_{ji}^\prime\eta_i^{\rm eff}}
\right.\right.\nonumber \\
&& \left.\left. - f(\beta_{ji}^\prime,\tilde{\tau}_\bot(\nu_{i+1}))
\sbkt{\frac{\nu_{i+1}}{\nu_j}}^{-\beta_{ji}^\prime\eta_i^{\rm eff}} \right\}
- \Gamma_j \right], 
\label{eq:rate4}
\end{eqnarray} 
where $\tilde{\tau}_\bot(\nu_{i+1}) =
\tau_\bot(\nu_i)(\nu_{i+1}/\nu_i)^ {-\eta_i^{\rm eff}}$, and a
function $f(a,x)$ is defined by 
\begin{eqnarray} 
\label{eq:f}
  f(a,x) &=& \frac{1}{a+1}\lbkt{\frac{\gamma(a,x)}{x^a} + e^{-x} - x
    E_1(x)}, \\
  &\rightarrow& \left\{\begin{array}{ll}
      \frac{1}{a} &  (x \rightarrow 0) , \\
      \frac{1}{a+1} \frac{\Gamma(a)}{x^a}   & (x \rightarrow \infty) .
     \end{array} \right. 
\label{eq:flim}
\end{eqnarray} 
using the incomplete gamma function (eq. [\ref{eq:gam}]) and the
exponential integral:
\begin{equation} 
  E_n(x) = x^{n-1} \int_x^\infty \frac{e^{-t}}{t^n} dt, ~~~~ x>0, ~~
  n=0,1,2...
\label{eq:expint}
\end{equation}  
The above results show that one can explicitly compute the
photoionization coefficients and heating rates, once given the column
densities perpendicular to the plane. In the optically thick limit
$\tau_\bot \rightarrow \infty$, the photoionization coefficients and
heating rates vary as $\tau_\bot^{-\beta}$ (see also
eqs~[\ref{eq:coeff7}][\ref{eq:rate7}] below).


\subsection{Simple example  ~~~ -- pure hydrogen gas --}

If the gas is dominated by a single species, say atomic hydrogen, the
expressions derived above can be reduced to even simpler forms as
presented below.  Such formulae are also quite useful in making
physical estimations in many astrophysical problems.

For the pure hydrogen gas, the effective indices $\eta_i^{\rm eff}$
can be all set equal to $\eta^{\rm eff}_{\rm HI} = \eta_{\rm HI}$, and
equations (\ref{eq:coeff4}) and (\ref{eq:rate4}) reduce to
\begin{eqnarray} 
\label{eq:coeff5}
  \Gamma_{\rm HI} &=& \frac{2\pi\sigma_{\rm HI}(\nu_{\rm HI})
    I_0(\nu_{\rm HI})}{h}
  \frac{f(\beta_{\rm HI},\tau_\bot(\nu_{\rm HI}) )}{\eta_{\rm HI}}, 
\\
{\cal H}_{\rm HI} &=& h \nu_{\rm HI} \left[\frac{2\pi\sigma_{\rm HI}
(\nu_{\rm HI}) I_0(\nu_{\rm HI})}{h} 
\frac{f(\beta_{\rm HI}^\prime,\tau_\bot(\nu_{\rm HI}) )}
{\eta_{\rm HI}} - \Gamma_{\rm HI}
\right] , 
\label{eq:rate5}
\end{eqnarray} 
with $\eta_{\rm HI}=3$, $\beta_{\rm HI}=(\alpha+3)/3$, and $\beta_{\rm
  HI}^\prime=(\alpha+2)/3$. Given a specific value of $\alpha$, these
  expressions are computed using equation (\ref{eq:flim}) in the
  optically thin and thick limits as
\begin{enumerate}
\item optically thin limit: $\tau_\bot(\nu_{\rm HI})\rightarrow 0$
\begin{eqnarray} 
\label{eq:coeff6}
  \Gamma_{\rm HI} &=& \left\{\begin{array}{ll}
      1.49\times 10^{-12} I_{21} \mbox{~~~s}^{-1} &  ~~~(\alpha =1) , \\
      7.47\times 10^{-13} I_{21} \mbox{~~~s}^{-1} &  ~~~(\alpha =5) ,
     \end{array} \right. \\
{\cal H}_{\rm HI} &=& \left\{\begin{array}{ll}
      1.08\times 10^{-23} I_{21} \mbox{~~~erg s}^{-1}&  (\alpha =1) , \\
      2.33\times 10^{-24} I_{21} \mbox{~~~erg s}^{-1}&  (\alpha =5) ,
     \end{array} \right. 
\label{eq:rate6}
\end{eqnarray} 
\item optically thick limit: $\tau_\bot(\nu_{\rm HI})\rightarrow \infty$ 
\begin{eqnarray} 
\label{eq:coeff7}
  \Gamma_{\rm HI} &=& \left\{\begin{array}{ll}
      7.62\times 10^{-13} I_{21} \tau_\bot^{-4/3}(\nu_{\rm HI})
 \mbox{~~~s}^{-1} &  ~~~(\alpha =1) , \\
      8.18\times 10^{-13} I_{21} \tau_\bot^{-8/3}(\nu_{\rm HI})
\mbox{~~~s}^{-1} &  ~~~(\alpha =5) ,
     \end{array} \right. \\
{\cal H}_{\rm HI} &=& \left\{\begin{array}{ll}
      2.17\times 10^{-23} I_{21} \tau_\bot^{-1}(\nu_{\rm HI})
\mbox{~~~erg s}^{-1}&  (\alpha =1) , \\
      1.55\times 10^{-23} I_{21} \tau_\bot^{-7/3}(\nu_{\rm HI})
\mbox{~~~erg s}^{-1}&  (\alpha =5) ,
     \end{array} \right. 
\label{eq:rate7}
\end{eqnarray} 
\end{enumerate}
where $I_{21}=I_0(\nu_{\rm HI}) /(10^{-21}\mbox{erg s$^{-1}$ cm$^{-2}$
str$^{-1}$ Hz$^{-1}$})$. Note that equations (\ref{eq:coeff5})--
(\ref{eq:rate7}) take account of the photons propagating from only
{\it one of the boundaries}. In particular, the expressions in the
optically thin limit (eqs~[\ref{eq:coeff6}] [\ref{eq:rate6}]) should be
multiplied by 2 to incorporate the photons coming from all directions.

From the above results, one can readily estimate the ionizing
structure of a medium exposed to the external photoionizing flux. We
restrict our attention to the gas of temperature $T \sim 10^4$K and
ignore the collisional ionization. Then the ionization balance is
expressed as
\begin{equation}
\Gamma_{\rm HI} X_{\rm HI} \sim \alpha_{\rm H} n_{\rm H} (1-X_{\rm HI})^2,
\label{eq:ioneql}
\end{equation}
where 
\begin{equation}
\alpha_{\rm H} \simeq 3.96 \times 10^{-13} T_4^{-0.7} 
\mbox{ ~cm$^3$ s$^{-1}$},  
\label{eq:hrec}
\end{equation}
is the hydrogen recombination rate to the ground level
(Spitzer 1978; Fukugita \& Kawasaki 1994), and $T_4 \equiv T/10^4
\mbox{K}$. Inserting equations (\ref{eq:coeff6}) and (\ref{eq:coeff7})
into (\ref{eq:ioneql}), we obtain
\begin{enumerate}
\item optically thin limit: $\tau_\bot(\nu_{\rm HI})\rightarrow 0$
\begin{equation} 
\label{eq:depth1}
 \frac{X_{\rm HI}}{(1-X_{\rm HI})^2} \sim \left\{\begin{array}{ll}
0.133 ~I_{21}^{-1} n_1 T_4^{-0.7} & ~~~(\alpha =1) , \\
0.265  ~I_{21}^{-1} n_1 T_4^{-0.7} & ~~~(\alpha =5) ,
\end{array} \right.
\end{equation} 
\item optically thick limit: $\tau_\bot(\nu_{\rm HI})\rightarrow \infty$
\begin{equation} 
\label{eq:depth2}
 \frac{X_{\rm HI}}{(1-X_{\rm HI})^2} \sim \left\{\begin{array}{ll}
     0.520  ~I_{21}^{-1} n_1   T_4^{-0.7}  
\tau_\bot^{4/3}(\nu_{\rm HI}) &  ~~~(\alpha =1) , \\
     0.484  ~I_{21}^{-1} n_1   T_4^{-0.7}  
\tau_\bot^{8/3}(\nu_{\rm HI})   &  ~~~(\alpha =5),  
\end{array} \right.
\end{equation} 
\end{enumerate}
where $n_1 \equiv n_{\rm H}/\mbox{cm}^{-3}$, and we have multiplied
equation (\ref{eq:coeff6}) by a factor 2 to incorporate the photons
coming from all directions. 

\bigskip
\bigskip
\section*{References}

\def\apjpap#1;#2;#3;#4; {\pp#1, {#2}, {#3}, #4}
\def\apjbook#1;#2;#3;#4; {\pp#1, {#2} (#3: #4)}
\def\apjppt#1;#2; {\pp#1, #2.}
\def\apjproc#1;#2;#3;#4;#5;#6; {\pp#1, {#2} #3, (#4: #5), #6}

\apjpap Anderson, S. F., Hogan, C. J., Williams, B. F.,\& Carswell, R. F. 
1999;AJ;117;56;
\apjpap Anninos, W. Y., \& Norman, M. L. 1994;ApJ;429;434;
\apjpap Babul, A., \& Rees, M. J. 1992;MNRAS;255;346; 
\apjpap Babul, A., \& Ferguson, H. C. 1996;ApJ;458;100; 
\apjpap Baljtlik, S., Duncan, R. C.,  \& Ostriker, J. P. 1988;ApJ;327;570;
\apjppt Barkana, R., \& Loeb, A. 1999;ApJ in press; 
\apjpap Barns J., \& Efstathiou, G. 1987;ApJ;319;575;
\apjpap Bechtold, J. 1994;ApJS;91;1;
\apjpap Berger, A. J., et al. 1998; Nature;394;248;
\apjpap Bertschinger, E. 1985;ApJS;58;39; 
\apjpap Bond, J. R., Szalay, A. S., \& Silk, J. 1988;ApJ;324;627;
\apjpap Bond, J. R., Cole, S., Efstathiou, G., \& Kaiser, N. 
1991;ApJ;379;440;
\apjbook Bowers, R. L., \& Wilson, J. R. 1991;Numerical Modeling in
Applied Physics and Astrophysics;Boston;Jones \& Bartlett;
\apjpap Cen, R. 1992;ApJS;78;341;
\apjpap Cen, R., \& Ostriker, J. P. 1992;ApJ;393;22;
\apjpap Cen, R., \& Ostriker, J. P. 1999;ApJ;514;1;
\apjpap Chengalur, J. N., \& Kanekar, N. 1999;MNRAS;302;L29;
\apjpap Chiba, M., \& Nath B. B. 1994;ApJ;436;618;
\apjpap Chieze, J.-P., Teyssier, R., \& Alimi, J.-M. 1997;ApJ;484;40;
\apjpap Cole, S., Aragon-Salamanca, A., Frenk, C. S., Navarro, J. F., \&
  Zepf, S. E. 1994;MNRAS;271;744; 
\apjpap Connolly, A. J., Szalay, A. S., Dickinson, M., Subbarao, M. U.,
 \& Brunner, R. J. 1997;ApJ;486;L11;  
\apjpap Cooke, A. J., Espey, B., \& Carswell, R. F. 1997;MNRAS;284;552;   
\apjpap Couchman, H. M. P. 1985;MNRAS;214;137;
\apjpap Couchman, H. M. P., \&   Rees, M. J. 1986;MNRAS;221;53;
\apjpap Efstathiou, G. 1992;MNRAS;256;43;
\apjpap Efstathiou, G. \& Jones, B. J. T. 1979;MNRAS;186;133;
\apjpap Eke, V. R., Cole, S., \& Frenk, C. S. 1996;MNRAS;282;263;
\apjpap Fukugita, M.,  \& Kawasaki, M. 1994;MNRAS; 269;563;
\apjpap Giallongo, E., Cristiani, S., D'Odorico, S., Fontana, A., \& 
  Savaglio, S. 1996;ApJ;466;46;
\apjpap Gnedin, N. Y., Ostriker, J. P. 1997;ApJ;486;581;  
\apjpap Gott, J. R. III 1975;ApJ;201;296;
\apjpap Gunn, J. E., \& Peterson, B. A. 1965;ApJ;142;1633; 
\apjpap Haiman, Z., \& Loeb, A. 1998;ApJ;503;505; 
\apjpap Haiman, Z., Thoul, A. A., \& Loeb, A. 1996;ApJ;464;523; 
\apjpap Haiman, Z., Rees, M. J., \& Loeb, A. 1996;ApJ;467;522;   
\apjpap Haiman, Z., Rees, M. J., \& Loeb, A. 1997;ApJ;476;458;  
\apjpap Hogan, C. J., Anderson, S. F. \& Rugers, M. H. 1997;AJ;113;1495; 
\apjpap Hughes, D., et al. 1998;Nature;398;241;
\apjpap Ikeuchi, S. 1986;Ap\&SS;118;509;
\apjpap Ikeuchi, S., Murakami, I., \& Rees, M. J. 1988;MNRAS;236;21P;
\apjpap Ikeuchi, S., Murakami, I., \& Rees, M. J. 1989;PASJ;41;1095;
\apjpap Kauffmann, G., White, S. D. M., \& Guiderdoni, B. 1993;MNRAS;264;201;
\apjpap  Kepner, J., Babul, A., \& Spergel, N. 1997;ApJ;487;61;
\apjpap Kitayama, T., \& Suto, Y. 1996a;MNRAS;280;638;
\apjpap Kitayama, T., \& Suto, Y. 1996b;ApJ;469;480;
\apjpap Kitayama, T., \& Suto, Y. 1997;ApJ;490;557;
\apjpap Lacey, C. G., \& Cole, S. 1993;MNRAS;262;627;
\apjpap Lilly, S. J., Le F\`{e}rve, O., Hammer, F., Crampton, D. 1996;ApJ;
460;L1;  
\apjpap Lane, W., Smette, A., Briggs, F., Rao, S., Turnshek, D., \& 
Meylan, G. 1998;ApJ;116;26;    
\apjpap Madau, P., Ferguson, H. C., Dickinson, M. E., Giavalisco M., 
Steidel, C. C., \& Fruchter, A. 1996;MNRAS;283;1388;
\apjpap Madau, P., Haardt, F., \& Rees, M. J. 1999;ApJ;514;648;
\apjpap Madau, P., Pozzetti, L., \& Dickinson, M. 1998;ApJ;498;106;
\apjpap Manrique, A., \& Salvador-Sole, E. 1996;ApJ;467;504; 
\apjpap Miralda-Escud\'{e} J., \& Ostriker, J. P. 1990;ApJ;350;1; 
\apjpap Murakami, I., \& Ikeuchi, S. 1990;PASJ;42;L11;  
\apjpap Murakami, I., \& Ikeuchi, S. 1993;ApJ;409;42;  
\apjpap Navarro, J. F., \& Steinmetz, M. 1997;ApJ;478;13;
\apjpap Okoshi, K. \& Ikeuchi, S. 1996;PASJ;48;441;
\apjbook Osterbrock, D. E. 1989;Astrophysics of Gaseous Nebulae and
 Active Galactic Nuclei;Mill Valley;University Science Books;
\apjbook Padmanabham, T. 1993;Structure Formation in the
 Universe;Cambridge;Cambridge Univ. Press; 
\apjpap Pettini, M., Kellog, M., Steidel, C. C., Dickinsonl, M., Adelberger, 
 K. L.,\& Giavalisco, M. 1998;ApJ;508;539; 
\apjpap Press, W. H., \& Schechter, P. 1974;ApJ;187;425;
\apjpap Quinn, T., Katz, N., \& Efstathiou, G. 1996;MNRAS;278;L49; 
\apjpap Rauch, M. 1998;ARA\&A;36;267;
\apjpap Rees, M. J. 1986;MNRAS;218;25P; 
\apjpap Reimers, D., K\"{o}hler, S., Wisotzki, L., Groote, D., 
Rodriguez-Pascual, P., \& Wamsteker, W. 1997;A\&A;327;890;
\apjbook Richtmyer, R., \& Morton, K. W. 1967; Difference Methods for
 Initial-Value Problems;New York;Interscience; 
\apjpap Sargent, W. L. W., Young, P. J., Boksenberg, A. \& Tytler, D. 
  1980;ApJS;42;41;  
\apjpap Sasaki, S. 1994;PASJ;46;427;
\apjpap Savaglio, S.; Cristiani, S., D'Odorico, S., Fontana, A.,
  Giallongo, E., \& Molaro, P. 1997;A\&A;318;347;  
\apjpap Smail, I., Ivison, R. J., \& Blain, A. W. 1997;ApJ;490;L5; 
\apjbook Spitzer, L. Jr. 1978;Physical Processes in the Intergalactic 
 Medium;Wiley;New York; 
\apjpap Spitzer, L. Jr., \& Hart, M. H. 1971;ApJ;166;483; 
\apjppt Susa, H., \& Umemura, M. 1999;ApJ submitted;
\apjpap Tajiri, Y., \& Umemura, M. 1998;ApJ;502;59;
\apjpap Thoul, A. A., \& Weinberg, D. H. 1995;ApJ;442;480; 
\apjpap Thoul, A. A., \& Weinberg, D. H. 1996;ApJ;465;608; 
\apjpap Totani, T., Yoshii, Y., \& Sato, K. 1997;ApJ;483;L75; 
\apjpap Umemura, M. 1993;ApJ;406;361;
\apjpap Umemura, M., \& Ikeuchi, S. 1984; Prog. Theor. Phys.;72;47;
\apjpap Umemura, M., \& Ikeuchi, S. 1985;ApJ;299;583;
\apjpap Vedel, H., Hellsten U., \& Sommer-Larsen J. 1994;MNRAS;1994;271;
\apjpap Viana, P. T. P., \& Liddle, A. R. 1996;MNRAS;281;323;
\apjpap Warren, M. S., Quinn, P. J., Salmon, J. K., \& Zurek, W. H. 
  1992;ApJ;399;405;
\apjpap Weinberg, D., Hernquist, L., \& Katz, N. 1997;ApJ;477;8;
\apjpap White, S. D. M., \& Frenk, C. S. 1991;ApJ;379;52;
\apjpap Zhang, Y., Anninos, P., \& Norman, M. L.
  1995;ApJ;453;L57;


\clearpage
\begin{table}
\caption{Characteristic redshifts of simulation runs}
\label{tab:zi}
\begin{center}
\begin{tabular}{l rrrrr} \\
\hline \hline \\[-6pt]
& $z_i$ & $z_{\rm ta0}$ & $z_{c0}$ & $z_{\rm ta}$ & $z_c$ 
   \\[4pt]\hline \\[-6pt]
low-redshift collapse& 40 & 5.4  & 3 & 1.3 & 0.5 \\  
middle-redshift collapse & 110 & 16 & 10 & 5.4 & 3 \\  
high-redshift collapse & 300 &  45 & 28 & 16 & 10 \\[4pt] 
\hline 
\end{tabular} 
\end{center}
\end{table}

\begin{table}
\caption{The critical impact parameter $p_{\rm crt}$ (see text for
definition) and the ratio of its square for the evolving $J_{21}$ and
$\alpha=1$}
\label{tab:cl1}
\begin{center}
\begin{tabular}{cccccc} \\
\hline \hline \\[-6pt]
  & &  \multicolumn{2}{c}{$V_c=32$ km s$^{-1}$} & 
\multicolumn{2}{c}{$V_c=50$ km s$^{-1}$} \\
 & species     & $p_{\rm crt}$ [kpc] 
 &  ratio of $p_{\rm crt}^2$ & $p_{\rm crt}$ [kpc] &  ratio of $p_{\rm crt}^2$ 
\\[4pt]\hline \\[-6pt]
      &HI & 10  & 1 & 10 & 1 \\
$z=3$ &HeI & 0.8 & 1/156 & 1 & 1/100\\
     &HeII & 20 & 4 & 40 & 16 \\[4pt]\hline \\[-6pt]
      & HI & 40  & 1 & 100 & 1 \\
$z=0$ &HeI & 10 & 1/16 & 10 & 1/100\\
      & HeII & 100 & 6.25 & 100 & 1 \\[4pt] 
\hline 
\end{tabular} 
\end{center}
\end{table}

\begin{table}
\caption{Same as Table \protect\ref{tab:cl1}\protect ~except for
 $J_{21}=1$ and $\alpha=5$.}
\label{tab:cl2}
\begin{center}
\begin{tabular}{cccccc} \\
\hline \hline \\[-6pt]
  & &  \multicolumn{2}{c}{$V_c=32$ km s$^{-1}$} & 
\multicolumn{2}{c}{$V_c=50$ km s$^{-1}$} \\
 & species     & $p_{\rm crt}$ [kpc] 
 &  ratio of $p_{\rm crt}^2$ & $p_{\rm crt}$ [kpc] &  
ratio of $p_{\rm crt}^2$ 
\\[4pt]\hline \\[-6pt]
      &HI & 20  & 1 & 30 & 1 \\
$z=3$ &HeI & 10 & 0.25 & 15 & 0.25\\
     &HeII & 100 & 25 & 150 & 25 \\[4pt]\hline \\[-6pt]
      & HI & 10  & 1 & 10 & 1 \\
$z=0$ &HeI & 7 & 0.49 & 7 & 0.49\\
      & HeII & 200 & 400 & 400 & 1600 \\[4pt] 
\hline 
\end{tabular} 
\end{center}
\end{table}

\begin{table}
\caption{Indices of column density distributions, $N^{\rm p}_j \propto
p^{-n}$ and $d{\cal N}/dN^{\rm p}_j \propto ({N^{\rm p}_j})^{-\beta}$}
\label{tab:cl3}
\begin{center}
\begin{tabular}{ccccc} \\
\hline \hline \\[-6pt]
  species & \multicolumn{2}{c}{$V_c=32$ km s$^{-1}$} & 
\multicolumn{2}{c}{$V_c=50$ km s$^{-1}$} \\
 $j$     & $n$ 
&  $\beta$ & $n$ &  $\beta$ 
\\[4pt]\hline \\[-6pt]
HI ($N^{\rm p}_{\rm HI} < 10^{16}$cm$^{-2}$) & 1.5 & 2.3 & 1.3 & 2.5 \\
HI ($N^{\rm p}_{\rm HI} > 10^{16}$cm$^{-2}$) & 3 & 1.7 & 24 & 1.1 \\
HeI & 3.7 & 1.5 & 18 & 1.1\\
HeII & 2 & 2 & 2.3 & 1.9 \\[4pt] 
\hline 
\end{tabular} 
\end{center}
\end{table}

\begin{table}
\caption{Relations among $N^{\rm p}_{\rm HI}$, $N^{\rm p}_{\rm HeI}$
and $N^{\rm p}_{\rm HeII}$ in cgs units}
\label{tab:cl4}
\begin{center}
\begin{tabular}{lll} \\
\hline \hline \\[-6pt] & \multicolumn{1}{c}{$N^{\rm
  p}_{\rm HI}< 10^{16}$cm$^{-2}$} &
  \multicolumn{1}{c}{$N^{\rm p}_{\rm HI}> 10^{16}$cm$^{-2}$} 
 \\[4pt]\hline \\[-6pt] $J_{21}$ evolving, $\alpha=1$ 
& $N^{\rm p}_{\rm HeI}\sim 10^{11} (N^{\rm p}_{\rm HI}/10^{14})^{1.5}$ 
& $N^{\rm p}_{\rm HeI} \sim 10^{-1} N^{\rm p}_{\rm HI}$ \\
& $N^{\rm p}_{\rm HeII}\sim 10 N^{\rm p}_{\rm HI}$ 
& $N^{\rm p}_{\rm HeII} \sim 10^{18}$ 
\\[4pt]\hline \\[-6pt] $J_{21}=1$, $\alpha=5$ 
& $N^{\rm p}_{\rm HeI}\sim 10^{13} (N^{\rm p}_{\rm HI}/10^{14})^{1.5}$ 
& $N^{\rm p}_{\rm HeI} \sim 10 N^{\rm p}_{\rm HI}$ \\
& $N^{\rm p}_{\rm HeII}\sim 10^3 N^{\rm p}_{\rm HI}$ 
& $N^{\rm p}_{\rm HeII} \sim 10^{18}$ 
\\[4pt]\hline
\end{tabular} 
\end{center}
\end{table}

\begin{table}
\caption{Photoionization cross section parameters in equation
(\protect\ref{eq:cross}\protect)}
\label{tab:cross}
\begin{center}
\begin{tabular}{cccc} \\
\hline \hline \\[-6pt]
species & ionization energy & amplitude & index \\
  $i$     &    $h \nu_i$ [eV] & $\sigma_i(\nu_i)$ [cm$^2$] & $\eta_i$ 
\\[4pt]\hline \\[-6pt]
HI (1)& 13.6  & $6.30 \times 10^{-18}$  & 3.0 \\
HeI (2)& 24.6 & $7.83 \times 10^{-18}$  & 2.05 \\
HeII (3)
& 54.4 & $1.58 \times 10^{-18}$  & 3.0 \\[4pt] 
\hline 
\end{tabular} 
\end{center}
\end{table}

\end{document}